\documentclass[aps,pre,reprint,superscriptaddress,10pt]{revtex4-2}
\usepackage{amsmath, amssymb, graphicx, color, braket,float}
\usepackage{bbold}
\usepackage{verbatim}

\begin{document}

\title{Tunable reentrant Kondo effect in quantum dots coupled to metal-superconducting hybrid reservoirs}

\author{Peter Zalom}
\email{zalomp@fzu.cz}
\affiliation{Institute of Physics, Czech Academy of Sciences, Na Slovance 2, CZ-18221 Praha 8, Czech Republic}

\author{Tom\'a\v{s} Novotn\'y}
\email{tno@karlov.mff.cuni.cz}
\affiliation{Department of Condensed Matter Physics, Faculty of Mathematics and Physics, Charles University, Ke Karlovu 5, CZ-12116  Praha 2, Czech Republic}

\date{\today}

\begin{abstract}
	We elaborate on the recently introduced concept of reentrant Kondo effect in quantum impurities/dots coupled to hybrid metal-semiconductor contacts [G.~Diniz {\em et al.}, Phys. Rev. B {\bf 101}, 125115 (2020)]. By noticing the equivalence of the originally suggested semiconducting arrangement to an analogous three-terminal quantum dot setup with a normal and two phase-biased superconducting leads introduced in our recent work [Phys. Rev. B {\bf 103}, 035419 (2021)], we put this effect into a new physical context, which enables us a fresh look on the problem. First, we identify the superconducting counterpart of the reentrant Kondo effect and, consequently, reveal its fragility with respect to an underlying doublet-singlet quantum phase transition induced by the particle-hole asymmetry of the reservoir densities of states. This is pertinent (even if previously unnoticed) also in the original as well as extended semiconducting setups, where it puts stringent conditions on the symmetry of the contact density of states. Furthermore, we analyze experimental feasibility of observing the reentrant Kondo effect in its superconducting realization concluding that even present day experiments might see the onset of the reentrant behavior, but fully developed Kondo features cannot be reached due to the required vast separation of energy/temperature scales.                
\end{abstract}

\maketitle

\section{Introduction \label{sec_intro} }

The Kondo effect is one of the most prominent many-particle phenomena
in which a localized magnetic moment is screened by the itinerant
electrons of its supporting host~\cite{Hewson97}. It has been found
to be important in various areas of solid state physics ranging from
the occurrence of zero-bias anomalies in transport measurements through
quantum dots \cite{Goldhaber98}, nanowires \cite{Nygard00}, and
single atoms or molecules \cite{Madhavan98,Yu04,Otte08a} to quantum-critical
phenomena like Mott metal-insulator transitions \cite{Nozieres05}
or possible applications in spintronics \cite{Bergmann15} and quantum
information transportation~\cite{Bayat12}. 

The generic microscopic model exhibiting the Kondo effect is the single impurity Anderson model \cite{Anderson-1961} with structureless tunnelling density of states of the metallic host/lead(s) whose solution has been well known already for decades \cite{Krishna-1980a,Hewson97,Bulla-Rev-2008}. Richer physics is found in cases of structured lead density of states (DOS) close to the Fermi energy, let it be a pseudogap \cite{Logan-2014} or zero-energy peaks, like van Hove singularities \cite{Bulla-Rev-2008, ZitkoAlen-2016}. Even less standard is the situation with a gapped tunneling density of states (TDOS) around the Fermi energy \cite{Takegahara-1992,Takegahara-1993, Saso-1992, Ogura-1993,Cruz-1995, Yu-1996, Chen-1998,Galpin-PRB2008,Galpin-EPJB2008,Moca-2010} which exhibits a quantum phase transition of the many-body ground state character depending on the model parameters, in particular the half-filling (or not) of the impurity/QD level. This model is also accompanied with the presence of bound state(s) within the gap.        

Recently Diniz {\em et al.}\ \cite{Diniz-2020} studied a modified setup with a strongly coupled gapped semiconducting lead and simultaneously a weakly coupled metallic lead, where they predicted an interesting regime of two reentrant Kondo resonances upon decreasing the temperature. While the characteristic energy scale of the high-temperature resonance $T_{K1}$ is larger then the energy of the semiconducting gap, the low-temperature one is observed at subgap energy scale $T_{K2}$. Interestingly, $T_{K1}$ follows well the Haldane formula for the single impurity Anderson model (SIAM) with bare parameters $\Gamma_S$ (coupling strength to the semiconductor) and $U$ (local Coulomb interaction). However, the second reentrant Kondo peak shows a more complex behaviour which is missing a good physical understanding.

As a physical realization of their model, Diniz {\em et al.}\  \cite{Diniz-2020} proposed to study the reentrant Kondo effect with an armchair graphene nanoribbon (AGNR) as the semiconducting substrate. Its TDOS possesses a gap induced by the Rashba spin-orbit interactions \cite{Lenz-2013} which can be to some extent manipulated by an external electric field. 

On the other hand, we have recently shown that a three-terminal set-up consisting of a half-filled quantum dot (QD) coupled to two phase biased superconducting leads and one metallic lead can be mapped via unitary transformation onto an Anderson spin $1/2$ impurity coupled to a structured electronic reservoir that retains the gap but off-diagonal superconducting correlations vanish. Moreover, setting the phase difference between the two superconducting electrodes to $\varphi=\pi$, the resulting model after the mapping is completely equivalent to the set-up proposed by Diniz {\em et al.}\  \cite{Diniz-2020}. Consequently, the reentrant Kondo effect can be rephrased in a different physical context, which offers not only a new physical implementation of this novel phenomenon but, as elaborated in this paper, also allows to deepen the physical understanding of the previously not well-understood spectral properties. 

In particular, we show using the superconducting realization of the model, that not only the particle-hole asymmetry of the impurity level (i.e.\ away from half-filling) but also the particle-hole asymmetry of the TDOS (i.e.\ its asymmetry with respect to the Fermi level) lead towards partial or complete disturbance of Kondo correlations due to the underlying quantum phase transition. Thus, the possibility of physical realization of the reentrant Kondo effect critically hinges on the ability to achieve and maintain particle-hole symmetry both on the impurity as well as in the surrounding leads/host. As shown in this paper, it affects superconducting as well as semiconducting hosts equally and may significantly impede chances to observe the reentrant Kondo effect. We note in this regard that the particle-hole symmetry of the TDOS is by no means generic for conventional semiconductors and only rather specific materials, such as the proposed AGNR in Ref.~\cite{Diniz-2020}, may offer sufficient level of particle-hole symmetry to warrant the existence of the effect. 

Therefore, the superconducting counterpart of the semiconducting setup may be a suitable alternative for the successful realization of the effect. We analyze this option in more detail, point out physical differences between the two implementations and conclude that there is a realistic chance to observe incipient traces of the reentrant Kondo behavior even in present-day superconducting experimental setups. The fully developed reentrant Kondo features are, however, hindered by the large disparity (several orders of magnitude) of required energy/temperature scales. This issue is, in our view, equally relevant and problematic also in the semiconducting implementation of the setup.           

The structure of the paper is as follows. In Sec.~\ref{sec_micro} we introduce the microscopic model for the superconducting implementation of the reentrant Kondo effect and summarize the transformation to the equivalent semiconducting case, while in Sec.~\ref{sec_nrg} we briefly introduce the numerical renormalization group (NRG) procedure used for obtaining our results. These are then presented in the following section \ref{sec_results} --- after preliminary analysis of the model parameter space framing the interesting regime in Sec.~\ref{subsec_preliminary} we present detailed NRG results for half-filled impurity at $\varphi=\pi$ in Sec.~\ref{subsec_pi}, for general $\varphi$ in Sec.~\ref{subsec_nopi}, and eventually even away from half-filling in Sec.~\ref{subsec_fragile}. We briefly comment on relation to experiments in Sec.~\ref{subsec_experiments} and finally conclude in Sec.~\ref{sec_conclusions}. 

\section{Theory \label{sec_theory}}

\subsection{Microscopic formulation \label{sec_micro} }

The hybrid three-terminal setup consists of a mesoscopic system modeled as a usual Anderson magnetic impurity connected to one normal metallic and two superconducting electrodes. The superconducting electrodes follow the Bardeen-Cooper-Schrieffer (BCS) theory with one lead referred to as the left ($L$) and the other one as the right ($R$), see Fig.~1 in Ref.~\cite{Zalom-2021}. The total Hamiltonian of the system is then the sum of the dot Hamiltonian $H_{d}$, the Hamiltonian of the normal lead $H_N$, two BCS Hamiltonians for superconducting leads $H_L$ and $H_R$, and three tunneling Hamiltonians $H_{T,\alpha}$ with $\alpha \in \{N, L, R\}$ which connect each lead separately to the dot. The constituent Hamiltonians read as 
\begin{eqnarray}
	H_d
	&=&
	\sum_{\sigma} 
	\varepsilon_{d}
	d^{\dagger}_{\sigma}
	d^{\vphantom{\dagger}}_{ \sigma}
	+
	U
	d^{\dagger}_{\uparrow}
	d^{\vphantom{\dagger}}_{ \uparrow}
	d^{\dagger}_{\downarrow}
	d^{\vphantom{\dagger}}_{ \downarrow},
	\label{dotH}
	\\
	H_{\alpha}  
	&=&
	\sum_{\mathbf{k}\sigma} 
	\, \varepsilon_{\mathbf{k}\alpha}
	c^{\dagger}_{\alpha \mathbf{k} \sigma}
	c^{\vphantom{\dagger}}_{\alpha \mathbf{k} \sigma}
	\nonumber
	\\
	&-&
	\Delta_{\alpha}
	\sum_{\mathbf{k}}
	\left(e^{i\varphi_\alpha} 
	c^{\dagger}_{\alpha \mathbf{k} \uparrow} 
	c^{\dagger}_{\alpha -\mathbf{k} \downarrow}
	+
	\textit{H.c.}\right),
	\label{kineticH}
	\\
	H_{T,\alpha} 
	&=&
	\sum_{\mathbf{k} \sigma} \,
	\left(V^*_{\alpha\mathbf{k}} 
	c^{\dagger}_{\alpha\mathbf{k}\sigma}
	d^{\vphantom{\dagger}}_{\sigma} 
	+ 
	V_{\alpha\mathbf{k}} 
	d^{\dagger}_{\sigma}
	c^{\vphantom{\dagger}}_{\alpha\mathbf{k}\sigma}\right),
	\label{tunnelH}
\end{eqnarray}
where $c^{\dagger}_{\alpha \mathbf{k} \sigma}$ creates an electron of spin $\sigma \in \{\uparrow \downarrow \}$ and quasi-momentum $\mathbf{k}$ in the lead $\alpha$ while $c^{\vphantom{\dagger}}_{\alpha\mathbf{k}\sigma}$ annihilates it. In analogy, $d^{\dagger}_{\sigma}$ creates a dot electron of spin $\sigma$ while $d^{\vphantom{\dagger}}_{\sigma}$ annihilates it. The QD is characterized by the Coulomb repulsion $U$ and the level energy $\varepsilon_d$, which in the most general case is arbitrary, but we will  only consider $\varepsilon_d=-U/2$ here. The QD hybridizes with the leads via $V_{\alpha\mathbf{k}}$ and the gap parameter vanishes in the normal lead, thus $\Delta_N=0$. 

As shown in Refs.~\cite{Zalom-2021,Domanski-2017}, the impact of the metal-superconductor reservoir can completely be expressed via a non-diagonal hybridization self-energy $\Sigma^d(\omega^+)$ \footnote{$\omega^+$ emphasizes to which part of the complex $z$-plain the function belongs, i.e.~above the real axis of $z$.} when the basis of $d$ fields is employed like previously done in Eqs.~(\ref{dotH})-(\ref{tunnelH}). However, performing a unitary transformation $\mathbb{T}: d \rightarrow w$ of the following form
\begin{equation}
W\equiv \mathbb{T} D
=
\frac{1}{\sqrt{2}}
\left(
\sigma_x - \sigma_z
\right) D\ ,
\label{transf_T}
\end{equation}
with Nambu spinors
\begin{equation}
D^{\dagger} 
=
\left(
d^{\dagger}_{\uparrow},
d^{\vphantom{\dagger}}_{\downarrow}
\right),
\qquad
W^{\dagger} 
=
\left(
w^{\dagger}_{\uparrow},
w^{\vphantom{\dagger}}_{\downarrow}
\right)\ ,
\end{equation}
and $\sigma_i$, $i\in \{x,y,x\}$ being the Pauli matrices allows to diagonalize the hybridization self-energy of the lead(s) while simultaneously transforms Eq.~(\ref{dotH}) to
\begin{eqnarray}
H_{\mathrm{dot}}
&=&
\sum_{\sigma} 
-\frac{U}{2}
w^{\dagger}_{\sigma}
w^{\vphantom{\dagger}}_{ \sigma}
+
U
w^{\dagger}_{\uparrow}
w^{\vphantom{\dagger}}_{ \uparrow}
w^{\dagger}_{\downarrow}
w^{\vphantom{\dagger}}_{ \downarrow}\ .
\label{H_d}
\end{eqnarray}

The noninteracting retarded Green function $\mathbb{G}^w_{0\sigma}(\omega^+)$ in the new basis $w$ adopts then a simple scalar spin-unpolarized form  (see Eq.~(30) in Ref.~\cite{Zalom-2021})
\begin{equation}
G^w_{0\uparrow} (\omega^+)
= 
G^w_{0 \downarrow} (\omega^+)
= 
\frac{1}{\omega +U/2 - \Sigma^w(\omega^+) }
\label{G_0w}
\end{equation}
with the scalar hybridization self-energy $\Sigma^w(\omega^+)$ fulfilling the Kramers-Kronig relations, so it is fully sufficient to specify just its imaginary part $\Gamma^w(\omega)$ known as the hybridization function or TDOS. For the present metal-superconducting reservoir it reads
\begin{equation}
\Gamma^w_{m-\mathrm{BCS}}(\omega;\varphi)
=
\Gamma_N
+
\frac{\Gamma_S|\omega|\Theta(\omega^2-\Delta^2)}
{\sqrt{\omega^2-\Delta^2}}
\left
(1-\frac{\Delta }{\omega}\cos\frac{\varphi}{2}
\right),
\label{BCS_DOS_w}
\end{equation}
where subscript $m-\mathrm{BCS}$ denotes the specific reservoir consisting of one metallic lead and two phase-biased (by the phase difference $\varphi$) BCS leads with identical gap $\Delta$ (as typical in SQUID experiments) \cite{Zalom-2021}. Both superconducting leads are hybridized to the QD by the same strength $\Gamma_S/2$ which suffices also for the construction of solution to any asymmetric configurations \cite{Kadlecova-2017} and $\Gamma_N$ is the hybridization strength of the metallic electrode solely. 

Clearly, the out-of-the gap region ($|\omega|>\Delta$) inherits the characteristic BCS gap divergences from the superconducting leads while the in-gap-region ($|\omega|<\Delta$) is completely described just by the metallic electrode. The resulting non-zero TDOS around the Fermi energy is then essential for the development of the low temperature ($T \lesssim\Delta$) Kondo correlations after the Hubbard interaction term is added to the problem, which then requires application of non-perturbative techniques. 
The scalar nature of the transformed problem in the $w$ basis (for convenience termed from now on the scalar basis) allows for the deployment of the otherwise unfeasible standard NRG approach \cite{Zalom-2021}, since the resulting physical model is that of a standard Anderson spin $1/2$ impurity coupled to a structured electronic reservoir without any off-diagonal superconducting correlations. 

In this respect, choosing $\varphi=\pi$ in the TDOS (\ref{BCS_DOS_w}) leads to the following two important conclusions. First, only for such a choice does $\Gamma^w_{m-\mathrm{BCS}}(\omega;\varphi)$ become particle-hole symmetric and, second, it is identical with the hybridization function  
\begin{equation}
\Gamma_{m-\mathrm{semi}}(\omega)
=
\Gamma_N
+
\frac{\Gamma_S|\omega|\Theta(\omega^2-\Delta^2)}
{\sqrt{\omega^2-\Delta^2}}=\Gamma^w_{m-\mathrm{BCS}}(\omega;\varphi=\pi)
\label{DOS_Diniz}
\end{equation}
recently studied in Ref.~\cite{Diniz-2020}. Thus, it completely describes two different physical realization of the reservoir which for one case is composed of one metallic and one semi-conducting lead while in the other case it includes one normal and two phase-biased BCS leads. Although the reservoir is of completely different physical nature we clearly observe its equivalence to the superconducting hybrid reservoir at $\varphi=\pi$ when expressed in the scalar basis $w$, which ensures that normal spectral functions are the same for both problems. 

The (normal) spectral function of the original superconducting model, i.e. the physical spectral function expressed in the original $d$ basis, is found by the symmetrization of the spectral function in the $w$ basis (see Eq.~(34) in Ref.~\cite{Zalom-2021}) as 
\begin{equation}\label{symmetrization}
 A_n^d(\omega) = \frac{1}{2}\left[A^w(\omega)+A^w(-\omega)\right].
\end{equation}   
Once again, choosing $\varphi=\pi$ in Eq.~(\ref{BCS_DOS_w}) leads due to particle-hole symmetry to an exceptional scenario where 
\begin{equation}\label{equality}
A^d_n(\omega,\varphi=\pi)=A^w(\omega,\varphi=\pi)=A_{m-\mathrm{semi}}(\omega), 
\end{equation} 
which establishes the equivalence between the semiconducting  problem  of Ref.~\cite{Diniz-2020} and the superconducting at $\varphi=\pi$  even in terms of the normal spectral properties in their physical bases. 

Therefore, the reentrant Kondo scenario proposed in the hybrid semiconducting experimental setup in Ref.~\cite{Diniz-2020}  should be equally observable in a three-terminal hybrid superconducting setup tuned to the $\varphi=\pi$ phase difference by the magnetic flux in the SQUID. Furthermore, the tunability of the phase difference via the magnetic flux in the superconducting context enables an extension of the reentrant Kondo effect to (effectively) particle-hole asymmetric electronic reservoirs, whose study we further pursue here.

\subsection{NRG solution \label{sec_nrg} }

To solve for the half-filled QD coupled to the metal-superconducting reservoir at arbitrary parameter values its scalar representation in the $w$ basis has been employed in accord with Ref.~\cite{Zalom-2021}. The resulting formulation of the model as a spin $1/2$ coupled to a scalar spin-unpolarized but frequency-dependent TDOS is of one-channel nature, which was the key motivation beyond performing the unitary transformation \eqref{transf_T}. Otherwise, the apparent three-channel nature of the original superconducting model for $\varphi\neq 0, \pi$ is intractable with the conventional NRG implementations. The structured TDOS provides the necessary coefficients to construct the Wilson chain for the corresponding NRG calculation \cite{Bulla-1994, Bulla-Rev-2008}. On the other hand, we stress that the $\Gamma_N = 0$ case is in the $w$ basis unsolvable, since for a scalar {\em gapped} TDOS the corresponding NRG discretization still needs to be developed, which is currently an ongoing research. Yet, the strict $\Gamma_N = 0$ case corresponds to the purely superconducting two-terminal SCIAM whose conventional NRG solution is well known and available within, e.g., the NRG Ljubljana package.

The three-terminal $\Gamma_N \neq0$ case has been solved in the $w$ basis by utilizing NRG Ljubljana code \cite{Ljubljana-code} with intertwined $z$-discretization according to the scheme of \v{Z}itko et al. \cite{ZitkoPruschke-2009}, where the discretization parameter was always set as $\Lambda=2$ and $z= n/10$ with $n \in \{ 0, \ldots 10 \}$. Nevertheless, even with such a large $z$-discretization some numeric artifacts are clearly visible as oscillatory features in spectral functions, see for example Figs.~\ref{fig_asym}$(e)$ and $(f)$. To achieve spectral functions with discontinuities at the BCS gap edges, the so-called self-energy trick has been employed \cite{Bulla-1998}. We stress that we concentrate exclusively on the wide band limit with bandwidth set typically to $2B=10^4\,\Delta$ or higher. The corrections for finite size bands have been performed following Ref.~\cite{Zalom-2021}. 

In this paper we concentrate on normal spectral properties, which are related to experimentally accessible quantities in setups involving scanning tunneling microscopy (STM) and, from the theoretical point of view, provide the unifying link between semiconducting and the superconducting case at $\varphi=\pi$.

\section{Results \label{sec_results} }

\subsection{Preliminary parameter space analysis \label{subsec_preliminary} } 

As already argued, Eq.~(\ref{equality}) ensures that in terms of the normal spectral function the proposed semiconducting system of Ref.~\cite{Diniz-2020} is equal to the superconducting hybrid three-terminal set-up of Ref.~\cite{Zalom-2021} when the phase difference $\varphi=\pi$ is selected. The superconducting system is then known to be in the so-called $\pi$-like regime where the subgap normal spectrum can be interpreted in terms of two broadened Andreev bound states (ABS) and one central Kondo peak \cite{Zalom-2021}. The latter one can be now identified as the reentrant, i.e.~low-temperature, peak observed in Ref.~\cite{Diniz-2020} for the proposed semiconducting system.

In superconducting realizations, however, one may exert additional control over the subgap properties by tuning the phase difference $\varphi$, which is possible by changing the magnetic flux in SQUID configurations. In such a case, one may start with the phase drop $\varphi=\pi$ and gradually decrease it until the crossover region to the $0$-like phase at $\varphi=\varphi^*$ is reached. In this process, the width of the central reentrant Kondo peak follows the $\log T_{K2} \propto \cos^2(\varphi/2)$ law as long as $\Gamma_N \lesssim\Delta/5$ \cite{Zalom-2021}. The resulting $\varphi$-controlled enhancement of the width $T_{K2}$ of the central reentrant Kondo peak is thus of essential advantage in our proposed superconducting realization of the reentrant Kondo effect. However, we stress out that a further decrease below $\varphi^*$ induces in the system the $0$-like regime whose main feature is the complete  destruction of the central Kondo peak and thus to a complete disruption of the reentrant Kondo effect.

The crucial point here is that the underlying particle-hole asymmetry of the TDOS in the scalar basis $w$ has been identified in Ref.~\cite{Zalom-2021} as the source of such a subgap behaviour. The destruction of the central peak in the superconducting system for $\varphi$ below $\varphi^*$ is therefore linked to the presence of sufficient amount of charge fluctuations in the system \cite{Zalom-2021}. Evidently, the reentrant Kondo effect is thus inherently fragile with respect to the inclusion of the particle-hole asymmetry and this applies also to the semiconducting environments which stayed completely unnoticed in Ref.~\cite{Diniz-2020}. To this end, one may generalize the semiconducting system of Ref.~\cite{Diniz-2020} by including the particle-hole asymmetry according to Eq.~\eqref{BCS_DOS_w}. The results of Ref.~\cite{Zalom-2021} for the superconducting scenario require then just a minor reinterpretation due to the different underlying reference basis ($w$ instead of $d$)  as done in Secs.~\ref{subsec_nopi} and \ref{subsec_fragile}. Accounting for even more general and realistic forms of the particle-hole asymmetry in semiconducting TDOS is also possible but goes beyond the scope of present work. Nevertheless, also such more general cases are expected to still maintain the same qualitative picture.

The particle-hole asymmetric destabilization of the reentrant Kondo effect is notably stronger when $\Gamma_N \gtrsim \Delta$, as shown  in Sec.~\ref{subsec_fragile}. Combined with the large $U$ as chosen in Ref.~\cite{Diniz-2020} the influence of even small amount of particle-hole asymmetry is then detrimental to the reentrant Kondo effect. In addition, the extreme parameter choices in Ref.~\cite{Diniz-2020} require gaining control over 8 orders of magnitude of temperature (or, equivalently, frequency). Taking these complications into account we have thus devised a more robust strategy for the analysis in both the superconducting as well as semiconducting realizations of the reentrant effect. 

\begin{figure}[t]
	\includegraphics[width=1.00\columnwidth]{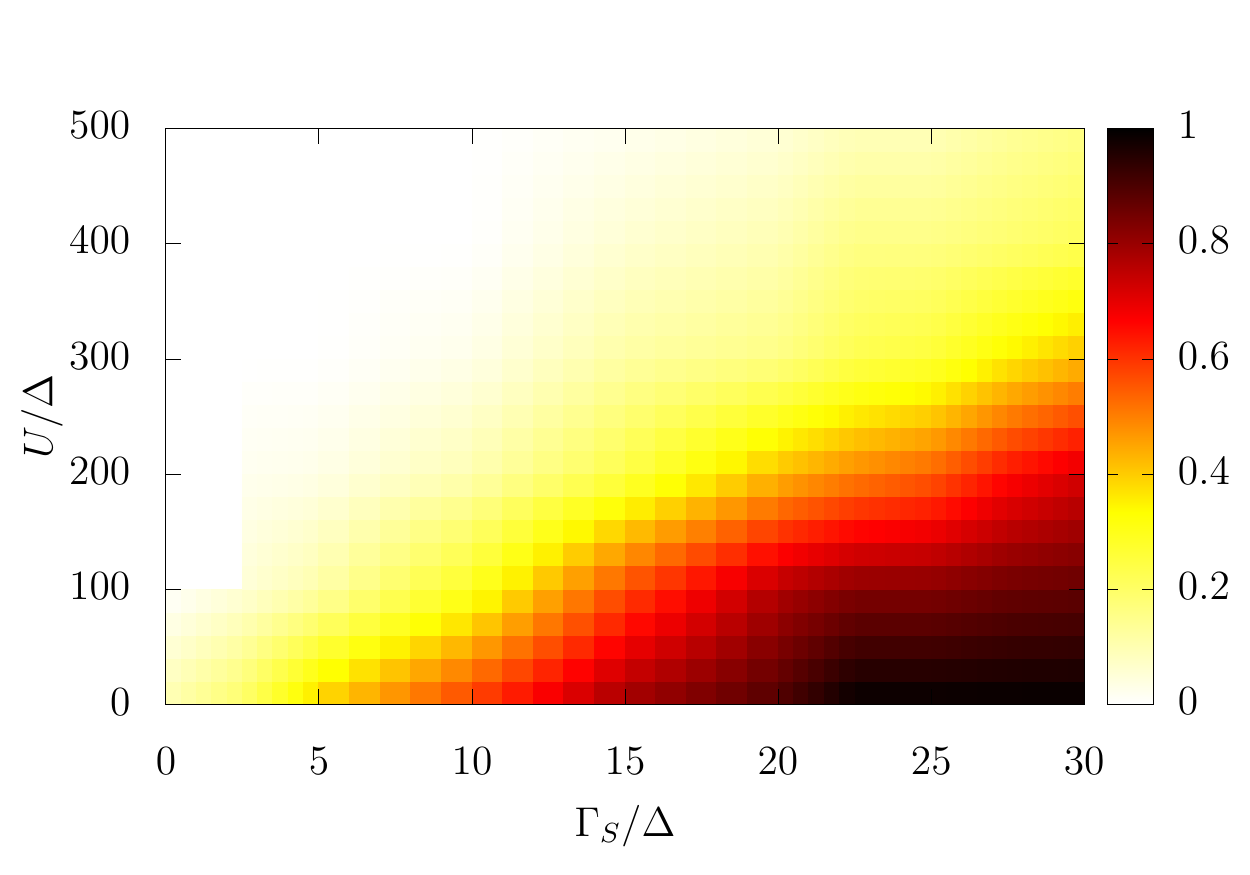}
	\caption{Compliance of the above-gap peak ($|\omega|\geq\Delta$) of the normal spectral function $A^d_n(\omega)$ with the Friedel sum rule in the parameter space of $U/\Delta$ and $\Gamma_S/\Delta$ on the scale between $0$ (above-gap peak not present) and $1$ (peak fulfilling the Friedel sum rule) for $\varphi=\pi$ and $\Gamma_N=\Delta/2$. NRG solution is given according to Sec.~\ref{sec_nrg}.
		\label{fig_preliminary}
	}
\end{figure}

We thus first set $\varphi=\pi$ to stay strictly in the particle-hole symmetric case and limit ourselves to the regime $\Gamma_N \lesssim \Delta$   and $U/\Delta \lesssim 500$, which is more stable against the inclusion of the particle-hole asymmetry and, beneficially, also lies within the reach of the current experiments. Under these limitations, we then conduct a preliminary scan in the parameter space of $U/\Delta$ and $U/\Gamma_S$ with respect to the properties of the high temperature (i.e.,~above gap) resonant features and at first completely omit the analysis of the low temperature (subgap) properties since the low- and high-temperature features largely decouple when $\Gamma_S \gg \Gamma_N$. To this end, we have thus set $\Gamma_N=\Delta/2$ in this preliminary analysis and assessed the compliance of the high temperature, above gap Kondo-like peak with the Friedel sum rule $A^d_n(\omega\gtrsim\Delta) \approx 1 /[\pi (\Gamma_N+\Gamma_S)]$.  The results are summarized in Fig.~\ref{fig_preliminary}. As the decision threshold for further analysis, we have set the level of compliance with the Friedel sum rule up to at least $75\%$, i.e.~$\pi (\Gamma_N +\Gamma_S) \max_{|\omega|\geq\Delta}A^d_n(\omega) > 0.75$.

Such preselection of the $U/\Delta$ and $U/\Gamma_S$ parameter values with respect to the fulfillment of the Friedel sum rule then ensures the existence of an above-gap peak in the normal spectral function, while the non-zero $\Gamma_N$ guarantees the existence of the reentrant peak for $|\omega|<\Delta$ which satisfies the Friedel sum rule  $A^d_n(\omega=0) \approx 1 /(\pi \Gamma_N)$. The necessary condition for the reentrant Kondo effect is then met and the suitable parameter regime can be easily read off from Fig.~\ref{fig_preliminary}.  Detailed spectral and NRG analysis of both peaks is then carried out in Sec.~\ref{subsec_pi}. The influence of the particle-hole asymmetry is then studied in Secs.~\ref{subsec_nopi} and \ref{subsec_fragile}.

Let us now look at the results of the preliminary analysis in more detail. According to Fig.~\ref{fig_preliminary}, the superconducting $\varphi=\pi$ three-terminal system fulfills the necessary conditions to support the reentrant Kondo effect in a portion of the parameter space. Choosing a moderate value of $U/\Delta \approx 100$ requires the parameter $\Gamma_S/\Delta \approx 30$ to detect an above-gap peak satisfying the Friedel sum rule with more than $90\%$ compliance. The resulting ratio of $U/\Gamma_S \approx 3$ (the impact of $\Gamma_N/\Delta=0.5$ can be still neglected), however, suggests that the above-gap peak may not fully correspond to the strongly correlated Kondo regime and requires thus a detailed NRG and spectral analysis. Let us stress that this problem is persistent to any $U/\Delta$ shown in Fig.~\ref{fig_preliminary} and can only be lifted at $\Gamma_N>\Delta$ but for the expense of fragility of the system against particle-hole asymmetry as shown in Sec.~\ref{subsec_fragile}. In what follows, we thus first concentrate on the regimes with $\Gamma_N \lesssim \Delta$ and intermediate values $U/\Delta \lesssim 100$ as they are more accessible in the experiments.

\subsection{Particle-hole symmetric TDOS ($\varphi=\pi$)}
\label{subsec_pi}

\begin{figure*}[t]
	\includegraphics[width=2.10\columnwidth]{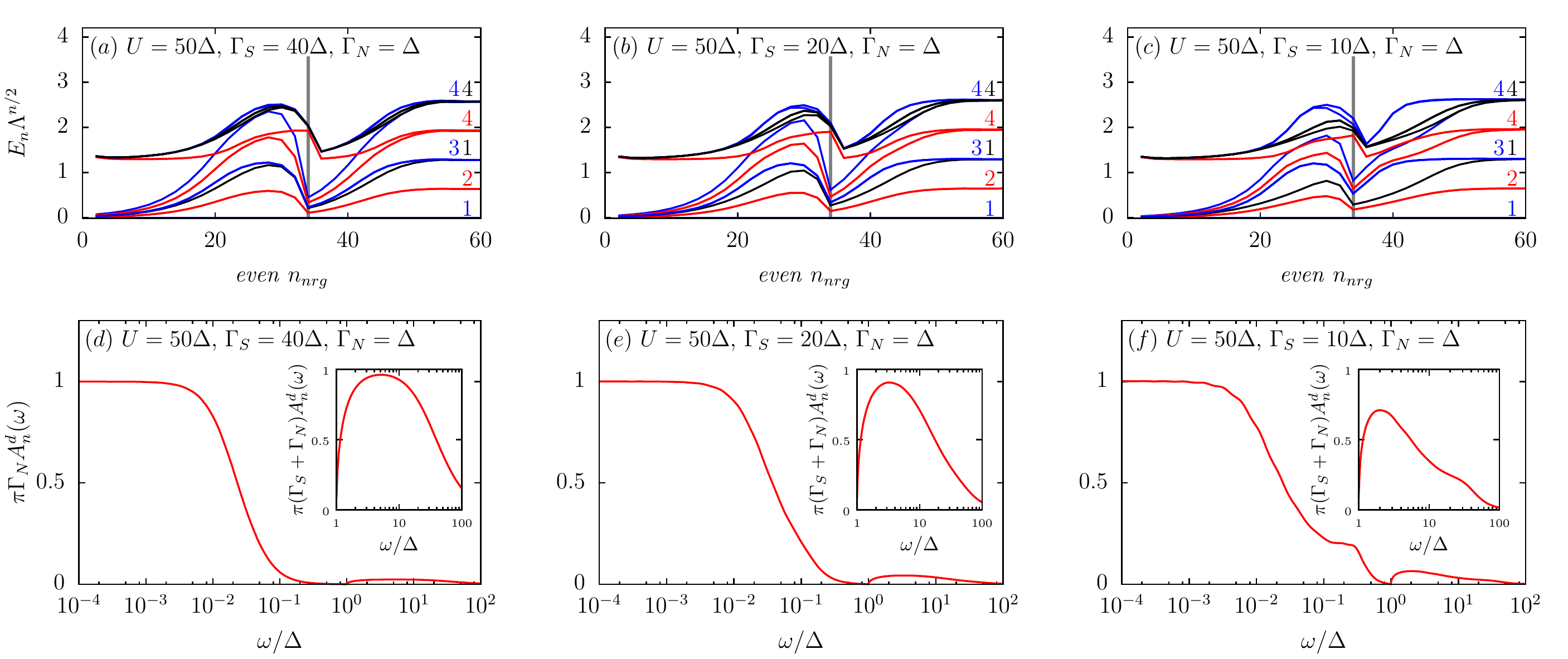}
	\caption{
		$(a)$-$(c)$
		NRG flow of rescaled energy eigenvalues  $E_n\Lambda^{n/2}$ at the $n^{\mathrm{th}}$ NRG iteration for the TDOS (\ref{DOS_Diniz}) calculated for selected values of $\Gamma_S$. Since several lines may overlap, we indicate their multiplicity at the end of the NRG flow by the attached numbers. The gray vertical line at $n_{\Delta}=34$ denotes the NRG iteration with the energy resolution $\Delta \approx B\Lambda^{-(n_{\Delta}-1)/2}$. It divides the NRG flow into high temperature ($n<34$) and low temperature parts ($n>34$). In the panel $(a)$ with $U/\Gamma_S=50/41\approx 1.22$ the high temperature NRG flow ($n>34$) just barely develops the strong coupling (SC) fixed point. The low temperature part clearly flows into the SC fixed point for $n\approx 50$. In the panel $(b)$ $U/\Gamma_S=2.5$, which causes the high temperature NRG flow to stop in the crossover region before the SC fixed point. The low temperature NRG flow develops SC fixed point for $n >54$. In the panel $(c)$ $U/\Gamma_S=5$ so that  the high temperature NRG flow misses the SC and valence-fluctuation fixed points entirely. The low temperature part flows evidently towards the SC point of the ordinary SIAM for $n>60$ which is however much smaller than $n\approx 90$  resulting from the NRG flow of the ordinary SIAM at the corresponding model parameters. 
		$(d)-(f)$
		Normal spectral functions \eqref{equality} for the parameters corresponding to the top panels. 
		All panels have been calculated using NRG Ljubljana in the wide band limit with $\Lambda=2$. In the SC fixed point, $E_n$ depend only on $\Lambda$ making the corresponding effective Hamiltonian identical for all panels.		
		\label{fig_sym}}
\end{figure*}

After the preselection performed in Sec.~\ref{subsec_preliminary} we can proceed with the further NRG and spectral analysis for the particle-hole symmetric case $\varphi=\pi$. Thus, Eq.~(\ref{equality}) holds and the results presented here apply both to the metal-semiconducting as well as to the metal-superconducting reservoirs. For a detailed discussion, we have selected $U/\Delta=50$ and three different values of $U/\Gamma_S$ which keep the Friedel sum rule fulfilled at least up to $75 \%$ for the above-gap peak. The corresponding NRG flows and the resulting normal spectral functions are shown in Figs.~\ref{fig_sym}$(a)$-$(c)$ and \ref{fig_sym}$(d)$-$(f)$, respectively. All results correspond to the wide band with $B= 10^{5}\Delta$.

The NRG flows show, that regardless of $U/\Gamma_S$ the singlet ground state and a degenerate pair of doublet excited states form almost instantly in the NRG flows while the higher-lying excited states form patterns characteristic for various NRG fixed points which are typically observed in the ordinary particle-hole symmetric SIAM. However, unlike in the ordinary particle-hole symmetric SIAM, the reentrant nature of the problem emerges as two consecutive NRG flows that connect at the $n_{\Delta}=34$ NRG iteration, where the energy resolution $B \Lambda^{-(n-1)/2}$ of the $n^{\mathrm{th}}$ NRG iteration reaches approximately $\Delta$ for the selected parameters ($\Lambda=2$). The NRG flow separates here into the high temperature ($n_{\mathrm{NRG}}<n_{\Delta}$) and low temperature ($n_{\mathrm{NRG}}>n_{\Delta}$) parts. Apparently, neither $\Gamma_S$ nor $\Gamma_N$ impact this crucial behavior which lies at the heart of the reentrant Kondo effect.

\begin{figure*}[ht]
	\includegraphics[width=2.00\columnwidth]{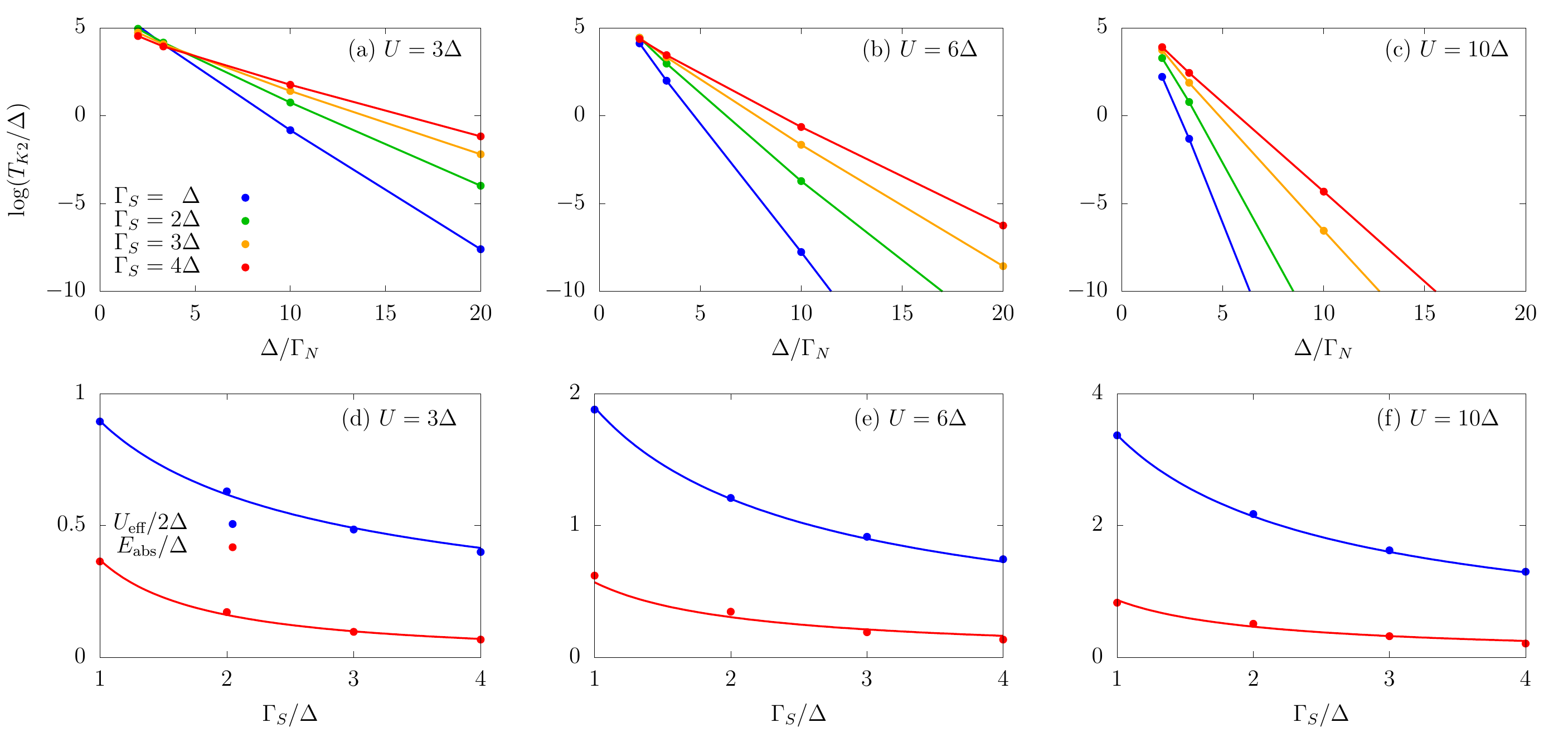}
	\caption{
		$(a)$-$(c)$ 
		Parametric $\Gamma_S$  plot of $\log(T_{K2}/\Delta)$ vs.~$\Delta/\Gamma_N$ for the hybrid superconducting system corresponding to the TDOS (\ref{DOS_Diniz}) and various values of $U$ at $\varphi=\pi$. Points represent the actual NRG data while the lines are just for visual guidance. Notice that the dependencies appear linear indicating a good compliance with the Haldane expression \eqref{Haldane} and allowing a reliable fitting of the proportionality factor $U_{\mathrm{eff}}$. $(d)$-$(f)$ Values of $U_{\mathrm{eff}}$ for the parameters corresponding to the top panels show non-trivial dependence on $U$ and $\Gamma_S$. The comparison to the energy of the ABS states $E_{\mathrm{abs}}$ when $\Gamma_N=0$ (SCIAM) shows that it correlates with $U_{\mathrm{eff}}$ but the two quantities do not coincide.
		\label{fig_Tk}}
\end{figure*}

These findings of the NRG flow analysis can be further deepened by inspecting the corresponding normal spectral functions $A^d_n(\omega)$ shown in Figs.~\ref{fig_sym}$(d)$-$(f)$ with insets highlighting their high frequency ($|\omega|>\Delta$) parts. To assess the scaling of the spectral functions we have used logarithmic frequency axes for positive frequencies $\omega$ both in the main panels as well as in the insets. Due to the particle-hole symmetry of the TDOS the normal spectral functions are symmetric around the Fermi energy and the negative frequency parts are therefore just mirror images of those presented in Fig.~\ref{fig_sym}.

As expected from our preliminary analysis, in all three cases the high frequency (temperature) peak shown in the insets conforms to the Friedel sum rule at least up to $75\%$ while the low frequency (temperature) central peak shows up at the Fermi energy and fulfills very precisely the sum rule  $A^d_n(\omega=0) = 1 /(\pi \Gamma_N)$. Additional features include the anticipated in-gap peaks (reported previously in Ref.~\cite{Diniz-2020}) which, however, at small ratios of $U/\Gamma_S$ [Figs.~\ref{fig_sym}$(d)$ and $(e)$] are indistinguishable from the central peak. Only at $U/\Gamma_S=5$ [Fig.~\ref{fig_sym}$(f)$] a good separation is observed. This suggests that in Figs.~\ref{fig_sym}$(d)$ and $(e)$ the strong coupling regime is not reached yet. The corresponding central peaks are then very broad and well understandable even at the level of the Hartree-Fock approximation. Thus, strictly speaking only the Fig.~\ref{fig_sym}$(f)$ shows a well developed low-temperature Kondo peak. On the other hand, as shown in the inset the high frequency (temperature) peak does {\em not} comply too well with the Friedel sum rule, it does not reach any plateau and thus strong charge fluctuations are mixed in. In the end, at moderate $U/\Delta$ both peaks cannot simultaneously reach the strongly coupled regime but they at least show clear signs of the incipient Kondo physics.

To reach strongly coupled regimes for both peaks, we could then try to relax our restriction on the intermediate values of $U/\Delta$ by pushing $\Delta/B$ to smaller values as performed in Ref.~\cite{Diniz-2020}. In the previously discussed cases one would, however, need to shift the corresponding $n_{\Delta}$ to at least $50$ which makes the gap $\Delta/B$ smaller by a factor of thousand (or the corresponding ratio of $U/\Delta$ thousand times bigger). Consequently, at intermediate values of $U/\Delta$ one is always facing the dilemma of either requiring extremely small and thus unfeasible energy resolutions in the experiment or to leave the strongly correlated regime. 

In other words, the reentrant Kondo regime described in Ref.~\cite{Diniz-2020} is realizable neither in semiconducting nor in the superconducting scenario if clear and unambiguous experimental signatures of strongly coupled Kondo regime are requested. However, resorting to a compromise, the reentrant nature of the problem is to be found in the normal spectral function always as a rudiment of the Kondo peak at large frequencies [see the inset of Fig.~\ref{fig_sym}$(f)$] and a well-formed central Kondo peak at $\omega\lesssim T_{K2}$. The advantage is that both peaks are experimentally well detectable as their signal is fairly strong and both appear within the range of two or three orders of magnitude of  frequency (temperature).

Let us now concentrate on the quantitative aspects of the central peak regarding the $\Gamma_S$ parameter. Decreasing $\Gamma_S$ by factor of two [compare Figs.~\ref{fig_sym}(d) and (e)] while keeping $\Gamma_N$ increases $T_{K2}$ of the central reentrant peak also roughly by factor of two, which is, however, to some extent unexpected as one would naively expect that it is only controlled by the hybridization strength of the normal lead $\Gamma_N$ and the Coulomb repulsion $U$. Instead, the enhancement of $T_{K2}$ then means that some effective $U_{\mathrm{eff}}$ and/or $\Gamma_{N,\mathrm{eff}}$ govern this behavior. However, according to Figs.~\ref{fig_sym}$(d)$-$(f)$ the Friedel sum rule is well fulfilled with $\pi \Gamma_N A^d(\omega)=1$ for $\omega \ll \Delta$. Consequently, the effective change of $\Gamma_N$ can be ruled out and an effective $U_{\mathrm{eff}}<U$ must govern the low temperature properties of the reentrant Kondo peak. 

In Ref.~\cite{Diniz-2020}, this behaviour was explained via the so-called effective Hubbard peaks located at $\omega\approx \pm U_{\mathrm{eff}}/2$. In the present case, similar in-gap peaks are clearly to be identified in Fig.~\ref{fig_sym}$(f)$ [while in Figs.~\ref{fig_sym}$(d)$-$(e)$ they are absorbed in the central peaks]. However, we stress out that in the superconducting scenario they correspond to the broadened ABS states which follows from a standard NRG calculation with disconnected metallic electrode. The resulting purely superconducting version of the model is known as SCIAM and has sharply located in-gap peaks at $E_{abs}$ which are known as the ABS states and for $\Gamma_N \lesssim \Delta/5$ coincide with the center of the broadened in-gap peaks of the normal spectral functions for the three-terminal case as demonstrated  already in Ref.~\cite{Zalom-2021}. In the superconducting, scenario the interpretation via effective Hubbard peaks can therefore be easily assessed. Moreover, due to equivalence \eqref{equality} the same applies by analogy also to the semiconducting case of Ref.~\cite{Diniz-2020}. However, the corresponding $\Gamma_N=0$ NRG calculation cannot be directly performed.

In this regard, we have therefore first extracted $T_{K2}$ as the half-width at half maximum (HWHM) of the corresponding central peak and fitted the data by the Haldane formula \cite{Hewson-1993}
\begin{equation}\label{Haldane}
\log(T_{K2}/\Delta) \propto -\pi U_{\mathrm{eff}}/(8\Gamma_N). 
\end{equation}
For this analysis, we have selected even much lower values of $U/\Delta$, in order to allow for a very wide range of other model parameters to be used in the corresponding NRG calculations. The results are shown in Fig.~\ref{fig_Tk} from which it follows that indeed $U_{\mathrm{eff}}<U$ as expected. However, comparing these values with $E_{\mathrm{ABS}}$ shows that $U_{\mathrm{eff}}/2$ is significantly larger, although the overall decrease with $\Gamma_S$ is correlated, see panels $(d)-(f)$ in Fig.~\ref{fig_Tk}. The position of the in-gap peaks $E_{abs}$ is thus crucial for diminishing $U_{\mathrm{eff}}$ compared to the bare $U$ but it does not truly coincide with the $U_{\mathrm{eff}}/2$ evaluated according to the Haldane formula. 

Consequently, the broadened analog of the ABS states cannot be simply described as an effective Hubbard peak. More complex screening processes are involved and also high energy features connected with the charge fluctuations according to the bare $U$ need to be taken into account. In the limit of infinitesimal $\Gamma_N$ (in practice, when $\Gamma_N$ is the smallest involved energy scale) this could be quantified by using the generalized Schrieffer-Wolff transformation in the spirit of Ref.~\cite[App. C]{Zitko-PhDthesis} and the resulting effective exchange coupling constant $J_{\mathrm{eff}}\equiv 8|V_{Nk_F}|^2/U_{\mathrm{eff}}$ expressed in the Lehmann representation could be evaluated via the NRG method of Ref.~\cite{Yang-2020}. This, however, goes far beyond the scope of the present work although it certainly constitutes an interesting open research problem.

\begin{figure*}[ht]
	\includegraphics[width=2.00\columnwidth]{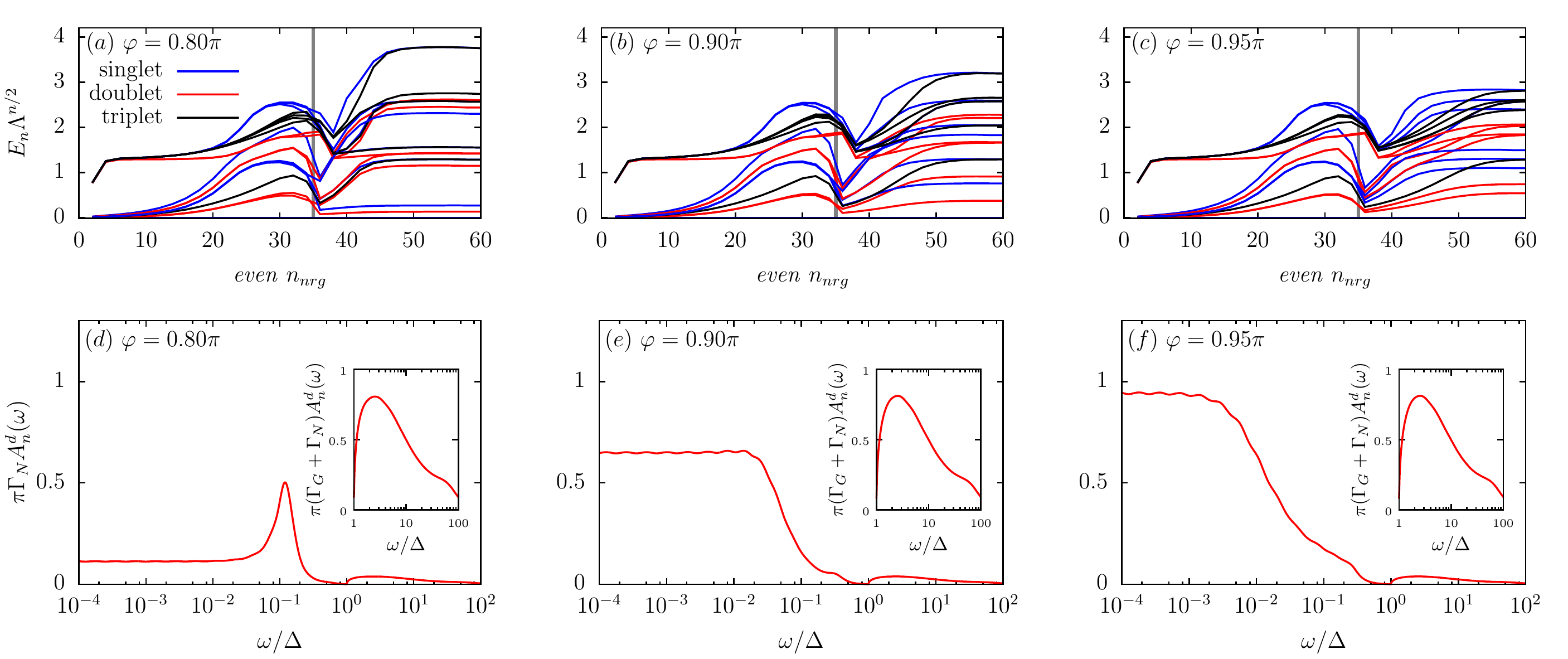}
	\caption{
		$(a)-(c)$ 
		NRG flow of $E_n$ for the TDOS functions~(\ref{BCS_DOS_w}) at $U=100\Delta$, $\Gamma_N=\Delta$, $U/\Gamma_S=5$, and selected values of $\varphi$. Notation is the same as in Fig.~\ref{fig_sym} including the gray vertical line at $n_{\Delta}=34$ denoting the NRG iteration with the energy resolution roughly $\Delta$. It divides the NRG flow into the high temperature ($n<34$) and the low temperature parts ($n>34$). In the panel $(a)$, $\varphi=0.8\pi$ and the superconducting hybrid system is in the $0$-like phase. Increasing $\varphi$ we reach the transition region at $\varphi \approx 0.90\pi$, see the panel $(b)$, and then the $\pi$-like phase for $\varphi=0.95\pi$ in the panel $(c)$. The high temperature NRG flow is largely independent of $\varphi$. On the contrary, the low temperature NRG flow shows various regimes typical of the particle-hole asymmetric SIAM.
		$(d)-(f)$ 
		Spectral functions of the particle-hole asymmetric semiconducting TDOS. Notation is the same as in Fig.~\ref{fig_sym}.
		$(g)-(h)$ 
		Normal spectral functions~\eqref{equality} for the particle-hole asymmetric superconducting TDOS~(\ref{BCS_DOS_w}). Parameter values correspond to the presented NRG flows from the top panels $(a)-(c)$. 	
		All panels have been  calculated using NRG Ljubljana in the wide band limit with $\Lambda=2$. Although the $z$-discretization scheme has been employed, noticeable numeric artifacts remain as unphysical oscillatory features of the low temperature plateau.	
		\label{fig_asym}}
\end{figure*}

\subsection{$\bf \varphi$-tunable reentrant Kondo effect}
\label{subsec_nopi}

The previous discussion at $\varphi=\pi$ relied on the properties of the superconducting and the corresponding semiconducting reservoirs which ensured the system to be in the $\pi$-like regime regardless of the remaining model parameters. Thus, the central reentrant Kondo peak was always present. Decreasing now the phase difference between the BCS leads to $\varphi < \pi$ introduces particle-hole asymmetry into the scalar TDOS (\ref{BCS_DOS_w}) which initially causes the central peak to become broader by following the $\log(T_{K2}) \propto \cos^2(\varphi/2)$ law as observed in Refs.~\cite{Zalom-2021,Domanski-2017}. 
However, in the crossover region $\varphi \approx \varphi^*$ the central peak vanishes due to the sufficient amount of charge fluctuations induced by the underlying particle-hole asymmetry of the TDOS --- an effect that is thoroughly discussed in Ref.~\cite{Zalom-2021} for $\Gamma_N \lesssim \Delta$ while additional new aspects for $\Gamma_N \gtrsim \Delta$ are discussed in Sec.~\ref{subsec_fragile}. 

Here, we exploit the enhancement of $T_{K2}$ in the $\pi$-like phase to exert control over the properties of the reentrant peak for $\Gamma_N \lesssim \Delta$. To this end, we have (based on the $\varphi=\pi$ discussion in Sec.~\ref{subsec_pi}) selected the ratio of $U/\Gamma_S=5$, where the high temperature peak is expected to show only some traces of Kondo correlations, however the low temperature peak can clearly be considered in the strongly coupled regime. To increase the $\varphi$ range of the $\pi$-like regime we have slightly increased the Coulomb repulsion to $U/\Delta=100$. The corresponding NRG flow is shown in Figs.~\ref{fig_asym}$(a)$-$(c)$ for three different phases $\varphi$. At $\varphi=0.95\pi$, see Fig.~\ref{fig_asym}$(c)$, the system is deep in the $\pi$-like regime. The $\varphi=0.90\pi$ case in Fig.~\ref{fig_asym}$(b)$ lies in the transition region between the $\pi$-like and the $0$-like phase and finally at $\varphi=0.80\pi$, we arrive at characteristic behavior deep in the $0$-like phase. 

Since $\Delta/B$ is kept the same as in Fig.~\ref{fig_sym}, the NRG flow divides again as before at $n_{\Delta}=34$ regardless of $\varphi$ into the high temperature ($n_{\mathrm{NRG}}<34$) and the low temperature part ($n_{\mathrm{NRG}}>34$). We first notice that the high temperature part of the NRG flow is almost unaffected by the increasing phase-induced particle-hole asymmetry of TDOS~(\ref{BCS_DOS_w}) in the scalar basis as expected since $\varphi$ controls predominantly a fairly narrow region of the TDOS around the gap edges. However, a non-trivial consequence of this finding is that by the phase-bias we can largely tune the $|\omega|<\Delta$ part of the spectral function while its  $|\omega|>\Delta$ counterpart corresponds mostly to the $\varphi=\pi$ case.

The low temperature part of the NRG flow now allows to assess the physical consequences of the phase-bias tuning on the low temperature peak. At $\varphi=\pi$ the analysis has shown that for the selected ratio $U/\Gamma_S$ the peak can be completely attributed to the strongly coupled Kondo regime. Decreasing $\varphi$ slightly to $0.95\pi$ causes the NRG eigenenergies to split symmetrically around approximately the positions they have occupied at $\varphi=\pi$ [compare with Fig.~\ref{fig_sym}$(c)$]. At $\varphi=0.95\pi$ the splitting is moderate and the system still remains in the regime of strong Kondo correlations, albeit competing with charge fluctuations. Decreasing the angle $\varphi$ further to $0.9\pi$ we reach the transition region between the $0$-like and the $\pi$-like phase and the splitting becomes larger, see Fig.~\ref{fig_asym}$(b)$, until finally at $\varphi=0.80\pi$ [Fig.~\ref{fig_asym}$(a)$] the extremely large splitting leads to a typical signature of the frozen-impurity fixed point of the particle-hole asymmetric SIAM. In both of these cases, the low temperature NRG flow clearly shows no signs of a strongly correlated Kondo regime which only underlines the necessity to tune the phase $\varphi$ in such a way that the metal-superconducting system remains clearly in the $\pi$-like phase.

\begin{figure*}[ht]
	\includegraphics[width=2.00\columnwidth]{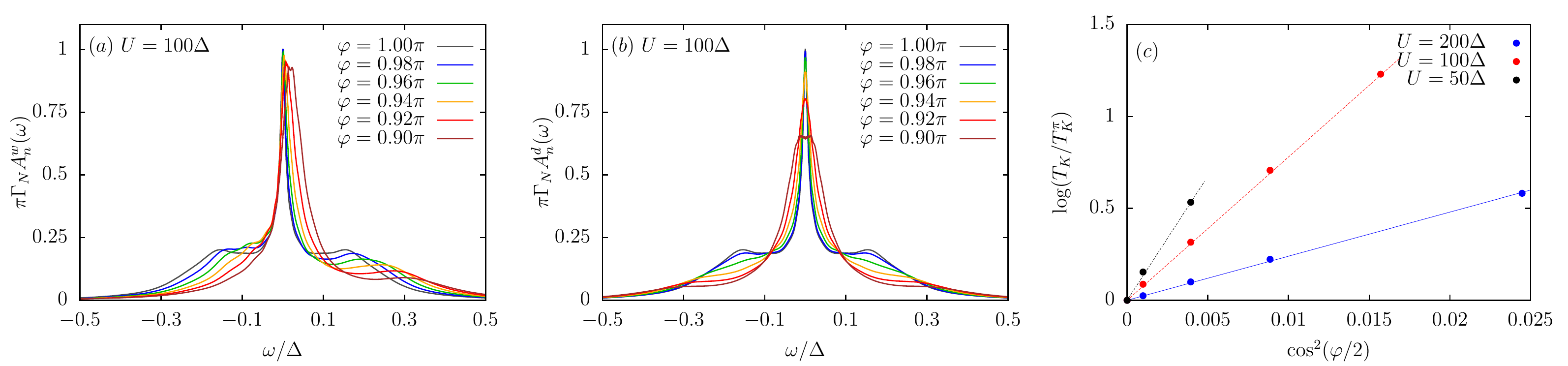}
	\caption{
		$(a)$		
		$\varphi$-evolution of the subgap part of the normal spectral function $A^w(\omega)$ in the scalar $w$ basis for TDOS (\ref{BCS_DOS_w}) and parameters $U=100\Delta$, $\Gamma_S=20\Delta$ and $\Gamma_N=\Delta$. Here, $\varphi$ merely parametrizes the particle-hole asymmetry of the TDOS and has no direct physical interpretation. We observe one  asymmetric central peak and two asymmetrically placed smaller peaks. All features show strong $\varphi$-dependence.
		$(b)$
		$\varphi$-evolution of the subgap part of the normal spectral function $A^d_n(\omega)$ in the natural basis $d$ with parameter values as in panel $(a)$. Spectral functions in panels $(a)$ and $(b)$ correspond to each other via symmetrization \eqref{symmetrization}. Correspondingly, in the superconducting scenario we observe one symmetric central peak and two pairs of smaller ABS peaks which merge together only at $\varphi=\pi$. 
		$(c)$
		Phase dependence of the Kondo temperature $T_{K2}$ for three selected values of $U$ at $\Gamma_N=\Delta$. $T_{K2}$ is 
		determined as the HWHM of the central Kondo-like peak observed in the $\pi$-like phase of the superconducting three-terminal set-up. $T_{K2}^{\pi}$ denotes $T_{K2}$ at $\varphi=\pi$. Points represent the NRG data while lines are fits in the corresponding $\pi$-like phase regions that terminate in the crossover region to the $0$-like region. Since $\Gamma_N = \Delta$, we observe not only a good agreement of the data with $\log T_{K2} \propto \cos^2(\varphi/2)$ but also enhancement of the $\pi$-like phase region as $\varphi^* \rightarrow 0$ with increasing $U$. 
		\label{fig_specs}}
\end{figure*}

The normal spectral functions $A^d_n(\omega)$ corresponding to the previously described NRG flows are shown in Figs.~\ref{fig_asym}$(d)$-$(f)$ on the logarithmic frequency scale. Concentrating first on the high frequency (temperature) part in the insets, we confirm the previous NRG findings about their independence of $\varphi$. However, as already concluded in the $\varphi=\pi$ case, neither of them does completely correspond to the strongly correlated regime with strong charge fluctuations being present which are connected with the Hubbard peak placed at approximately $U/2$ (visible as a shoulder of the discussed high-frequency peak). Additionally, the maximum value of the high-frequency peak complies with the Friedel sum rule only partially. 

Most crucially however, comparing the in-gap part of normal spectral functions in Fig.~\ref{fig_asym}$(e)$ and Fig.~\ref{fig_asym}$(f)$ one observes a significant increase of $T_{K2}$ when $\varphi$ is lowered toward $\varphi^*$. Noticeably, the behaviour is correlated with the overall lowering of the central peak. At $\varphi=0.8\pi$, in the $0$-like phase, we then observe only a small albeit non-zero spectral weight at the Fermi energy. Nevertheless, in the $\pi$-like phase the central peak is always well formed and becomes broader as $\varphi \rightarrow \varphi^*$. 
	
For a detailed analysis, we further study the $\varphi$-evolution of the subgap part of the normal spectral functions. Keeping $U$, $\Gamma_S$ and $\Gamma_N$ the same as in as in Fig.~\ref{fig_asym}, we set $\varphi$ to strictly stay in the $\pi$-like phase region. Using linear frequency scale, we present the subgap spectral functions $A^w(\omega)$ in the basis $w$ in Fig.\ref{fig_specs}$(a)$ while $A^d_n(\omega)$ in Fig.\ref{fig_specs}$(b)$ corresponds to the natural basis $d$. The resulting $T_{K2}$ values obtained as HWHM of $A^d_n(\omega)$ are then shown in Fig.~\ref{fig_specs}$(c)$. We stress that $A^d_n(\omega)$ in Fig.\ref{fig_specs}$(b)$ are obtained via symmetrization \eqref{symmetrization} of $A^w(\omega)$ in Fig.\ref{fig_specs}$(a)$ which are moreover relevant for QDs hybridized to metal-semiconductor reservoirs via TDOS \eqref{BCS_DOS_w}, but the discussion is kept for later.

For $U=100\Delta$, Fig.\ref{fig_specs}$(c)$ clearly establishes $\varphi$-dependent enhancement of $T_{K2}$ by a factor of roughly $4$ obtained at $\cos^2(\varphi/2) \approx 0.015$ when compared to the $\varphi=\pi$ case. However, we stress that by taking $\varphi \rightarrow \varphi^*$ the Kondo plateau diminishes which signals the weakening of the Kondo correlations \cite{Zalom-2021}. Furthermore, the enhancement of $T_{K2}$ also crucially depends on the extent of the $\pi$-like regime as shown in Fig.~\ref{fig_specs}$(c)$. For example, at $U=50\Delta$ the broadening of the central peak is more rapid as $\varphi \rightarrow \varphi^*$ but the transition to the $0$-like region occurs at much larger $\varphi^*$ [see the termination of the black line in Fig.~\ref{fig_specs}$(c)$ for $\cos^2(\varphi/2) \approx 0.005$]. Thus overall, the enhancement factor is smaller than in the case of $U=100\Delta$. At $U=200\Delta$, on the other hand, the $\pi$-like phase region becomes larger but the broadening is less rapid and when the crossover region at $\varphi \approx \varphi^*$ is reached [not shown in Fig.~\ref{fig_specs}$(c)$] the overall $T_{K2}$-enhancement is lower compared to the $U=100\Delta$ case. Moreover, at $U=200\Delta$ the values of $T_{K2}/\Delta$ become very small and thus experimentally hard to access. Thus beneficially, at moderate values of $U$, which are experimentally more reachable, the $\varphi$-tunning of $T_{K2}$ is more significant.
		
Let us now briefly analyze $A^w(\omega)$ obtained for the semiconducting realization with TDOS~(\ref{BCS_DOS_w}) as already presented in Fig.~\ref{fig_specs}$(a)$. They correspond via symmetrization \eqref{symmetrization} to the previously discussed spectral functions for the metal-superconductor case and require thus only an appropriate reinterpretation when basis $w$ used to describe the corresponding metal-semiconductor system. In particular, for semiconducting systems, $\varphi$ plays a role of the particle-hole asymmetry parameter, with $\varphi=\pi$ corresponding to the fully symmetric case. Starting here, we observe two in-gap bound states placed symmetrically around the Fermi energy and one central peak. Decreasing $\varphi$ induces then particle-hole asymmetry which deforms the central peak, i. e. it becomes asymmetric. Moreover, it shifts together with both in-gap peaks toward positive infrequences. We stress that due to the missing symmetrization \eqref{symmetrization}, we always observe only two broadened in-gap bound states unlike in the superconducting scenario. For selected parameters, we eventually observe at $\varphi^* \approx 0.9\pi$ that the particle-hole asymmetry is large enough to induce charge fluctuations that shift the negative frequency peak across the Fermi energy where it then also merges with the rudiment of the central peak. Besides of the missing symmetrization, the situation is reminiscent of the $0$-$\pi$ transition well known from the $\Gamma_N=0$ superconducting systems. 

However, due to the finite $\Gamma_N$ the crossing cannot be accompanied by the disappearance of the second in-gap peak which is now strongly expelled to the gap edge since the ground state of the system remains always of Kondo singlet nature as already explained in Ref.~\cite{Zalom-2021}. Although there are striking similarities between the semiconducting $\Gamma_N=0$ case and its superconducting counterpart, the later has the benefit of viable NRG solution scheme \cite{Satori-1992}. Consequently, underlying QPT is well known and can be used to understand the $\Gamma_N \neq 0$ scenario. However, for the semiconducting case no NRG solution exists so far because a hard spectral gap results in ill defined coefficients of the Wilson chain when standard logarithmic discretization of NRG is used. The understanding of the underlying QPT is thus hindered. Nevertheless, transformation \eqref{transf_T} presupposes that the NRG eigenspectrum of SCIAM is in one-to-one correspondence to the semiconducting $\Gamma_N=0$ realization. Such a conclusion is however only to be made for a very specific TDOS \eqref{BCS_DOS_w} with BCS-like singularities on the gap edges. An interesting question is thus to which extent can the specific TDOS \eqref{BCS_DOS_w} be varied while the $0$-$\pi$ transition related behaviour is preserved.  

Altogether, for $\Gamma_N \lesssim \Delta$ the phase-bias $\varphi$ allows to enhance $T_{K2}$ almost independently of the high temperature (frequency) region which is beneficial for successful experimental observation of the reentrant Kondo effect. However, for $\varphi < \pi$ we simultaneously increasingly include charge fluctuations to the character of the low temperature peak which poses a compromise to a large extent. This implies that sharp unambiguous character of the strongly coupled Kondo regime is hard to achieve if experimental restrictions are taken into account. These findings equally well apply to the superconducting as well as to the semiconducting cases. However, the superconducting realization allows for a straightforward phase-bias control of these phenomena in SQUID experiments, while a way to engineer tunable particle-hole asymmetric semiconducting TDOS of the form (\ref{BCS_DOS_w}) is to our best knowledge unknown.

\subsection{Particle-hole-asymmetry-induced instability of the reentrant Kondo effect \label{subsec_fragile} }

\begin{figure*}[ht]
	\includegraphics[width=2.00\columnwidth]{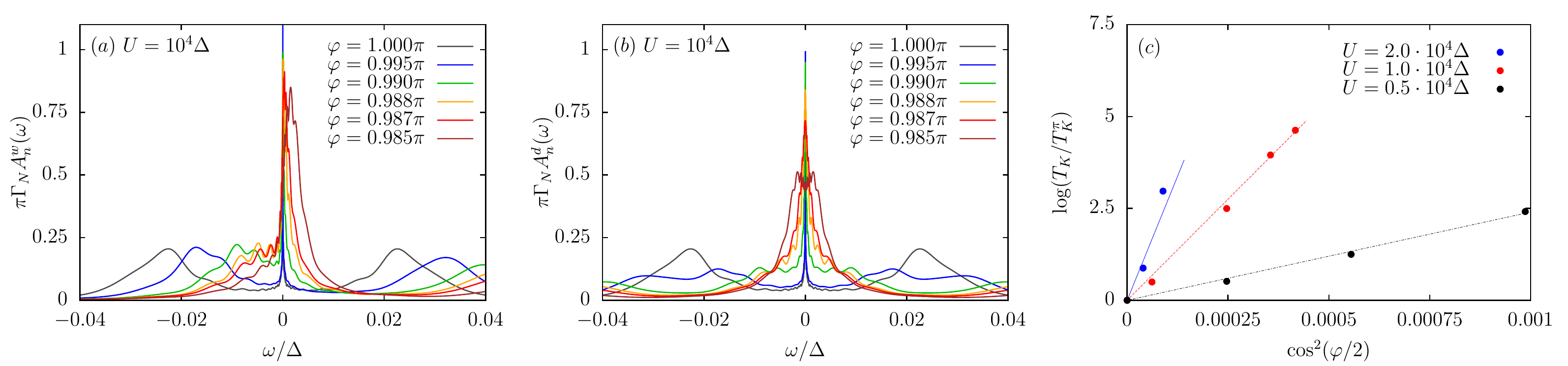}
	\caption{
		$(a)$		
		$\varphi$-evolution of the subgap normal spectral function $A^w(\omega)$ in the scalar $w$ basis for $U=10^4\Delta$, $U/\Gamma_S=10$ and $\Gamma_N=2.5\Delta$ for system with TDOS (\ref{BCS_DOS_w}).  
		$(b)$
		$\varphi$-evolution of the subgap normal spectral function $A^d_n(\omega)$ in the natural basis $d$ with parameter values as in panel $(a)$. Spectral functions in panels $(a)$ and $(b)$ correspond to each other via symmetrization \eqref{symmetrization}. Correspondingly, in the superconducting scenario we observe one symmetric central peak and two pairs of smaller ABS peaks which merge together only at $\varphi=\pi$. 
		$(c)$
		Phase dependence of the Kondo temperature $T_{K2}$ for $\Gamma_N \gtrsim \Delta$ shows significant differences when compared to the results of Fig.~\ref{fig_specs}$(c)$. HWHM is determined as previously, $T_{K2}^{\pi}$ denotes $T_{K2}$ at $\varphi=\pi$. Points represent the NRG data while lines are fits in the corresponding $\pi$-like phase regions that terminate in the crossover region to the $0$-like region. In all three cases, $\log T_{K2} \propto \cos^2(\varphi/2)$ but $\varphi^* \rightarrow \pi$ with increasing $U$ which is in stark contrast to the opposite shift in the $\Gamma_N \lesssim \Delta$ case.
		\label{fig_specs_diniz}}
\end{figure*}
	
A successful experimental realization of the reentrant Kondo effect critically hinges on the proper selection of parameters and as argued above is crucially affected by the particle-hole asymmetry of the TDOS. Up to now, we have selected $\Gamma_N \lesssim \Delta$ causing the behaviour of the three-terminal superconducting system to be well understandable from the properties of the underlying SCIAM. The influence of the normal electrode is then similar to the effect of temperature \cite{Domanski-2017} with the exception of the existence of the low temperature Kondo peak at $T_{K2}$. Moreover, as shown in the previous section as well as in Ref.~\cite{Zalom-2021} for $\Gamma_N \lesssim \Delta$ the phase bias tuning towards $\varphi^*$ increases $T_{K2}$ in the $\pi$-like region which itself becomes larger with increasing Coulomb interaction $U$, which is once again in analogy to the results for SCIAM.

On the other hand, in Ref.~\cite{Diniz-2020}, the authors chose the $\Gamma_N \gtrsim \Delta$ regime which has some remarkable but up to now unscrutinized features regarding the influence of the particle-hole asymmetry. Moreover, this regime is additionally combined with extremely large values of $U/\Delta$ and $\Gamma_S/\Delta$. We first note that such parameters cause the two Kondo scales of the reentrant Kondo effect to span over {\em eight orders of magnitude}, which is an experimentally unfeasible scenario. Additionally as unnoticed in Ref.~\cite{Diniz-2020}, combination with the $\Gamma_N\gtrsim\Delta$ choice makes the experimental observation of the reentrant Kondo effect even more elusive as the low temperature Kondo peak becomes extremely fragile against inclusion of the particle-hole asymmetry into the TDOS, an effect to be discussed here. 

First, we maintained the ratio $U/\Gamma_S=10$ and set $\Gamma_N=2.5 \Delta$ in accord with the parameters of Ref.~\cite{Diniz-2020}. For a better $\varphi$-resolution we slightly lowered the Coulomb repulsion on the QD to $U=10^4\Delta$ as compared to $U=5 \cdot 10^4 \Delta$ in Ref.~\cite{Diniz-2020}. As previously, the results in the $w$ basis, see Fig.~\ref{fig_specs_diniz}$(a)$, apply directly to the semiconducting case while those in the $d$ basis, see Fig.~\ref{fig_specs_diniz}$(b)$, describe the corresponding superconducting case. Once again, for the semiconducting case $\varphi$ needs to be understood merely as a parametrization of the particle-hole asymmetry of the given TDOS. Unlike for $\Gamma_N \lesssim \Delta$ however, we now observe a rapid splitting of the central peak that occurs at $\varphi^* \approx 0.985\pi$ which in light of the results presented in Fig.~\ref{fig_specs}$(c)$ seems unexpected for such a large value of $U$.

Lowering $\varphi$ further across $\varphi^*$ causes the transition into the $0$-like regime, where the central Kondo peak disappears leaving only the high temperature Kondo peak present. As a result, the reentrant Kondo effect becomes disrupted. The same applies also to the superconducting counterpart with symmetrized normal spectral functions shown in Fig.\ref{fig_specs_diniz} $(b)$. The crucial observation is then inferred from the $T_{K2}$ behaviour with respect to the increasing value of $U$. As shown in Fig.~\ref{fig_Tk}$(c)$ the $\Gamma_N \gtrsim \Delta$ regime causes $\varphi \rightarrow \pi$ with increasing $U$ which is in the opposite direction as expected from the underlying SCIAM or as observed for $\Gamma_N \lesssim \Delta$. Consequently, at $U/\Delta=5 \cdot 10^4$ selected in Ref.~\cite{Diniz-2020} even a relatively small disturbance of the particle-hole symmetry, i. e. $\varphi \sim 0.999\pi$, is sufficient to induce the transition into the $0$-like regime which is then detrimental for a successful experimental realization of the reentrant Kondo effect. Consequently, at $\Gamma_N\gtrsim\Delta$ the effect of $T_{K2}$ enhancement via $\varphi$ tuning becomes not only irrelevant in both the superconducting as well as the semiconducting scenario but shows that even a small introduction of particle-hole asymmetry into the semiconducting TDOS is detrimental to the reentrant Kondo effect.

We also conjecture that generalizing the semiconducting TDOS functions from the special superconducting-like \eqref{BCS_DOS_w} to a more general and realistic ones, would lead to the same qualitative effects. Such calculations are viable within our NRG scheme for $\Gamma_N > 0$. Moreover, such calculations could also shed light to the currently unfeasible $\Gamma_N=0$ NRG calculations which could clarify the presence of QPTs in QDS coupled to semiconductors.

\subsection{Relation to the experiments}\label{subsec_experiments}

Analogously to Ref.~\cite{Diniz-2020} we focused our study on the evaluation of the spectral function of the model. The somewhat nontrivial question is what is the relation of such equilibrium spectral functions to the experimentally measured quantities such as the differential conductance in transport measurements. In general, one should in fact study a corresponding nonequilibrium problem, which is a much harder task. However, if the coupling to the normal lead is much less than the coupling to the super/semi-conductor ($\Gamma_N\ll\Gamma_S$), which is the case both in our as well as Diniz's \cite{Diniz-2020} study, one can assume that the quantum dot/impurity is primarily equilibrated with the super/semi-conducting host and the normal electrode only serves as a weakly coupled probe as in, e.g., STM experiments. 

This physically intuitive notion can be formalized by an explicit calculation, see for example Ref.~\cite[Sec. 8.4.1]{Flensberg-book} for a simplest linear-response approach, which shows that for small $\Gamma_N$ the tunneling spectroscopy of the differential conductance probes the spectral function of the system {\em without the normal lead} at the frequency given by the applied bias voltage (this result is true also for the combination of one normal and the other superconducting leads as is obvious from the presented derivation). This result is sufficient for the interpretation of our above-gap results where the effect of finite $\Gamma_N$ is indeed negligible. Situation is more tricky with the interpretation of the subgap results which are inherently dominated by the finiteness of $\Gamma_N$ --- the very existence of the reentrant Kondo peak as well as the broadening of the ABS spectral peaks, which constitute the core of our findings within the gap, do not exist without finite, although possibly small $\Gamma_N$. 

Thus, for the interpretation of the subgap results we need a finer approach. In the superconducting case, it was argued in Ref.~\cite[Sec. V]{Domanski-2017} that the differential conductance is determined by the Andreev transmittance which can be expressed in terms of the anomalous component of the (equilibrium) Green function. As we showed in Ref.~\cite{Zalom-2021}, Eqs.~(33) and (35), this component can be straightforwardly expressed in terms of the spectral function in the $w$ basis $A^w(\omega)$ and, moreover, the width of the reentrant Kondo peak measured via the Andreev conductance can be directly related to that from the normal component of the spectral function studied here. This then makes a direct connection of our results and the potentially measured ones. The experimental interpretation of subgap results in the semiconducting scenario of Ref.~\cite{Diniz-2020} is somewhat unclear to us, but we expect that the tunneling differential conductance will be given in a reasonable approximation going beyond the linear response of Ref.~\cite{Flensberg-book} by the spectral function of the composite system (i.e., both with the semiconducting host and the normal lead), although it is only a conjecture.  

As for the existence of potentially relevant experiments, the appropriate setups have already been successfully realized by the Copenhagen group \cite{Jellinggaard-2016, Whiticar-2021}. The first experiment \cite{Jellinggaard-2016} consists of a correlated quantum dot attached to a single superconducting lead and weakly probed by a normal electrode. The lack of control over the superconducting phase difference is fixed in the newer setup \cite{Whiticar-2021} with a SQUID loop, which thus fully realizes our superconducting version of the studied model. Yet, no observation of the reentrant Kondo effect has been reported so far, which is most likely given by different parameter regimes in which these setups were operated. Given the rather high tunability in fabrication and consequent operation of such devices, we are convinced that the reentrant Kondo regime should be reachable in a dedicated experimental search for it.

\section{Conclusions}\label{sec_conclusions}

Using the results of Ref.~\cite{Zalom-2021} we have established equivalence (\ref{equality}) between the metal-semiconducting QD system studied in Ref.~\cite{Diniz-2020} and the three-terminal QD set-up of Ref.~\cite{Zalom-2021}. The two systems are shown to exactly match in their normal spectral properties when $\varphi=\pi$ is selected in the later realization. Consequently, both represent potential platforms for the observation of the reentrant Kondo effect as described in Ref.~\cite{Diniz-2020}. 

Although the normal spectral function of the $\varphi=\pi$ case is exactly the same as in the metal-semiconducting QD system of Ref.~\cite{Diniz-2020} we have subjected the superconducting analog to a thorough scrutiny regarding experimental realizability in the regime $\Gamma_N \lesssim \Delta$ which was not studied in Ref.~\cite{Diniz-2020}. The resulting findings of the preliminary analysis in Sec.~\ref{subsec_preliminary} followed by subsequent NRG and spectral analysis in Sec.~\ref{subsec_pi} identify stringent bounds to the observation of the reentrant Kondo effect at experimentally accessible moderate values of $U/\Delta \lesssim 100$. Summing up, at such parameter values both the subgap as well as the above gap peak may be well observed. However, charge fluctuations are strongly mixed in, so strictly speaking unambiguous strongly correlated Kondo regimes cannot be reached. Nevertheless, the reentrant nature is clearly to be identified from the corresponding tunneling specroscopy measurements of the normal spectral functions, while the corresponding theoretical analysis can prove that relevant contributions of Kondo correlations can be recovered. 

Another consequence of equality \eqref{equality} is the one-to-one correspondence of two physically different systems with the superconducting realization being very well understood both qualitatively as well as quantitatively. Consequently, we notice that the subgap peaks identified as effective Hubbard-like in-gap peaks in Ref.~\cite{Diniz-2020} are just broadened analogs of the ABS states observed in the superconducting realizations. As indicated in Fig.~\ref{fig_Tk}, they contribute together with the charge fluctuations due to the bare $U$ to the formation of the subgap Kondo correlations which are then consequently governed by an effective $U_{\mathrm{eff}}$.

Moreover, the superconducting case allows to fine tune the reentrant Kondo peak when $\Gamma_N \lesssim \Delta$ via changing the superconducting phase difference $\varphi$ by the magnetic flux in SQUID experiments. Decreasing $\varphi$ toward $\varphi^*$ one remains in the $\pi$-like phase where the subgap Kondo peak broadens according to the $\log(T_{K2}) \propto \cos^2(\varphi/2)$ law. Although in the resulting reentrant Kondo peak charge fluctuations are enhanced, phase-bias-tuning shifts $T_{K2}$ significantly up  (almost by one order of magnitude) in the experimentally relevant parameter ranges which raises the chances for its successful experimental observation.

Additionally, in the underlying scalar description of the superconducting three terminal set-up $\varphi$ tunes the particle-hole symmetry of the TDOS. Then, at $\varphi=\pi$, the completely particle-hole symmetric scalar TDOS ensures that the system is in the $\pi$-like regime which guarantees the presence of the reentrant subgap Kondo peak. However, at some critical particle-hole asymmetry, reached at $\varphi=\varphi^*$, the subgap Kondo peak is completely destroyed. Consequently, introducing particle-hole asymmetry into the scalar gapped TDOS can be detrimental for the existence of the reentrant Kondo peak. In the underlying scalar description of the superconducting three terminal set-up the system is physically interpreted as the magnetic impurity subjected to the interaction with a scalar gapped TDOS. Consequently, the description completely corresponds to the physics of a magnetic impurity coupled to a particle-hole (a)symmetric semiconducting environment augmented by a normal electrode. We thus show that particle-hole asymmetry, as can be expected in realistic semiconductors, drastically reduces the available parameter space for $U$ in which the reentrant Kondo peak can be realized.

Taking together, in this paper we did not only identify a novel platform to support the observation of the reentrant Kondo effect but we also narrowed down the requirements for its successful observation both in the semiconducting as well as in superconducting systems. In both realizations, the particle-hole symmetry of the corresponding scalar TDOS should be preserved as much as possible and the Coulomb repulsions needs to be kept at moderate levels to warrant the existence of the reentrant Kondo peak whose observability in the superconducting realization should be within the reach of current experimental setups.

\begin{acknowledgements}
	We thank Dr.~Panch Ram for numerical assistance in the beginning of this project. 
	This work was supported by Grant  No.~19-13525S of the Czech Science Foundation (TN), 
	by grant INTER-COST LTC19045 (PZ), 
	by the COST Action NANOCOHYBRI (CA16218) (TN),
	the National Science Centre (NCN, Poland) via Grant No.~UMO-2017/27/B/ST3/01911 (TN) and
	by The Ministry of Education, Youth and Sports from the Large Infrastructures for Research, 
	Experimental Development and Innovations project 
	``IT4Innovations National Supercomputing Center – LM2015070'' and project 
	``e-Infrastruktura CZ" (e-INFRA LM2018140).
\end{acknowledgements}


\begin{thebibliography}{59}%
	\makeatletter
	\providecommand \@ifxundefined [1]{%
		\@ifx{#1\undefined}
	}%
	\providecommand \@ifnum [1]{%
		\ifnum #1\expandafter \@firstoftwo
		\else \expandafter \@secondoftwo
		\fi
	}%
	\providecommand \@ifx [1]{%
		\ifx #1\expandafter \@firstoftwo
		\else \expandafter \@secondoftwo
		\fi
	}%
	\providecommand \natexlab [1]{#1}%
	\providecommand \enquote  [1]{``#1''}%
	\providecommand \bibnamefont  [1]{#1}%
	\providecommand \bibfnamefont [1]{#1}%
	\providecommand \citenamefont [1]{#1}%
	\providecommand \href@noop [0]{\@secondoftwo}%
	\providecommand \href [0]{\begingroup \@sanitize@url \@href}%
	\providecommand \@href[1]{\@@startlink{#1}\@@href}%
	\providecommand \@@href[1]{\endgroup#1\@@endlink}%
	\providecommand \@sanitize@url [0]{\catcode `\\12\catcode `\$12\catcode
		`\&12\catcode `\#12\catcode `\^12\catcode `\_12\catcode `\%12\relax}%
	\providecommand \@@startlink[1]{}%
	\providecommand \@@endlink[0]{}%
	\providecommand \url  [0]{\begingroup\@sanitize@url \@url }%
	\providecommand \@url [1]{\endgroup\@href {#1}{\urlprefix }}%
	\providecommand \urlprefix  [0]{URL }%
	\providecommand \Eprint [0]{\href }%
	\providecommand \doibase [0]{http://dx.doi.org/}%
	\providecommand \selectlanguage [0]{\@gobble}%
	\providecommand \bibinfo  [0]{\@secondoftwo}%
	\providecommand \bibfield  [0]{\@secondoftwo}%
	\providecommand \translation [1]{[#1]}%
	\providecommand \BibitemOpen [0]{}%
	\providecommand \bibitemStop [0]{}%
	\providecommand \bibitemNoStop [0]{.\EOS\space}%
	\providecommand \EOS [0]{\spacefactor3000\relax}%
	\providecommand \BibitemShut  [1]{\csname bibitem#1\endcsname}%
	\let\auto@bib@innerbib\@empty
	\bibitem [{\citenamefont {Hewson}(1997)}]{Hewson97}%
	\BibitemOpen
	\bibfield  {author} {\bibinfo {author} {\bibfnamefont {A.~C.}\ \bibnamefont
			{Hewson}},\ }\href@noop {} {\emph {\bibinfo {title} {Kondo Problem to Heavy
				Fermions, The}}}\ (\bibinfo  {publisher} {Cambridge University Press},\
	\bibinfo {address} {Cambridge},\ \bibinfo {year} {1997})\BibitemShut
	{NoStop}%
	\bibitem [{\citenamefont {Goldhaber-Gordon}\ \emph {et~al.}(1998)\citenamefont
		{Goldhaber-Gordon}, \citenamefont {Shtrikman}, \citenamefont {Mahalu},
		\citenamefont {Abusch-Magder}, \citenamefont {Meirav},\ and\ \citenamefont
		{Kastner}}]{Goldhaber98}%
	\BibitemOpen
	\bibfield  {author} {\bibinfo {author} {\bibfnamefont {D.}~\bibnamefont
			{Goldhaber-Gordon}}, \bibinfo {author} {\bibfnamefont {H.}~\bibnamefont
			{Shtrikman}}, \bibinfo {author} {\bibfnamefont {D.}~\bibnamefont {Mahalu}},
		\bibinfo {author} {\bibfnamefont {D.}~\bibnamefont {Abusch-Magder}}, \bibinfo
		{author} {\bibfnamefont {U.}~\bibnamefont {Meirav}}, \ and\ \bibinfo {author}
		{\bibfnamefont {M.~A.}\ \bibnamefont {Kastner}},\ }\href@noop {} {\bibfield
		{journal} {\bibinfo  {journal} {Nature}\ }\textbf {\bibinfo {volume} {391}},\
		\bibinfo {pages} {156} (\bibinfo {year} {1998})}\BibitemShut {NoStop}%
	\bibitem [{\citenamefont {Nyg\aa{}rd}\ \emph {et~al.}(2000)\citenamefont
		{Nyg\aa{}rd}, \citenamefont {Cobden},\ and\ \citenamefont
		{Lindelof}}]{Nygard00}%
	\BibitemOpen
	\bibfield  {author} {\bibinfo {author} {\bibfnamefont {J.}~\bibnamefont
			{Nyg\aa{}rd}}, \bibinfo {author} {\bibfnamefont {D.~H.}\ \bibnamefont
			{Cobden}}, \ and\ \bibinfo {author} {\bibfnamefont {P.~E.}\ \bibnamefont
			{Lindelof}},\ }\href@noop {} {\bibfield  {journal} {\bibinfo  {journal}
			{Nature}\ }\textbf {\bibinfo {volume} {408}},\ \bibinfo {pages} {342}
		(\bibinfo {year} {2000})}\BibitemShut {NoStop}%
	\bibitem [{\citenamefont {Madhavan}\ \emph {et~al.}(1998)\citenamefont
		{Madhavan}, \citenamefont {Chen}, \citenamefont {Jamneala}, \citenamefont
		{Crommie},\ and\ \citenamefont {Wingreen}}]{Madhavan98}%
	\BibitemOpen
	\bibfield  {author} {\bibinfo {author} {\bibfnamefont {V.}~\bibnamefont
			{Madhavan}}, \bibinfo {author} {\bibfnamefont {W.}~\bibnamefont {Chen}},
		\bibinfo {author} {\bibfnamefont {T.}~\bibnamefont {Jamneala}}, \bibinfo
		{author} {\bibfnamefont {M.~F.}\ \bibnamefont {Crommie}}, \ and\ \bibinfo
		{author} {\bibfnamefont {N.~S.}\ \bibnamefont {Wingreen}},\ }\href@noop {}
	{\bibfield  {journal} {\bibinfo  {journal} {Science}\ }\textbf {\bibinfo
			{volume} {280}},\ \bibinfo {pages} {567} (\bibinfo {year}
		{1998})}\BibitemShut {NoStop}%
	\bibitem [{\citenamefont {Yu}\ and\ \citenamefont {Natelson}(2004)}]{Yu04}%
	\BibitemOpen
	\bibfield  {author} {\bibinfo {author} {\bibfnamefont {L.~H.}\ \bibnamefont
			{Yu}}\ and\ \bibinfo {author} {\bibfnamefont {D.}~\bibnamefont {Natelson}},\
	}\href@noop {} {\bibfield  {journal} {\bibinfo  {journal} {Nano Lett.}\
	}\textbf {\bibinfo {volume} {4}},\ \bibinfo {pages} {79} (\bibinfo {year}
	{2004})}\BibitemShut {NoStop}%
\bibitem [{\citenamefont {Otte}\ \emph {et~al.}(2008)\citenamefont {Otte},
	\citenamefont {Ternes}, \citenamefont {Loth}, \citenamefont {von Bergmann},
	\citenamefont {Brune}, \citenamefont {Lutz}, \citenamefont {Hirjibehedin},\
	and\ \citenamefont {Heinrich}}]{Otte08a}%
\BibitemOpen
\bibfield  {author} {\bibinfo {author} {\bibfnamefont {A.~F.}\ \bibnamefont
		{Otte}}, \bibinfo {author} {\bibfnamefont {M.}~\bibnamefont {Ternes}},
	\bibinfo {author} {\bibfnamefont {S.}~\bibnamefont {Loth}}, \bibinfo {author}
	{\bibfnamefont {K.}~\bibnamefont {von Bergmann}}, \bibinfo {author}
	{\bibfnamefont {H.}~\bibnamefont {Brune}}, \bibinfo {author} {\bibfnamefont
		{C.~P.}\ \bibnamefont {Lutz}}, \bibinfo {author} {\bibfnamefont {C.~F.}\
		\bibnamefont {Hirjibehedin}}, \ and\ \bibinfo {author} {\bibfnamefont
		{A.~J.}\ \bibnamefont {Heinrich}},\ }\href@noop {} {\bibfield  {journal}
	{\bibinfo  {journal} {Nature Physics}\ }\textbf {\bibinfo {volume} {4}},\
	\bibinfo {pages} {847} (\bibinfo {year} {2008})}\BibitemShut {NoStop}%
\bibitem [{\citenamefont {Nozi\`{e}res}(2005)}]{Nozieres05}%
\BibitemOpen
\bibfield  {author} {\bibinfo {author} {\bibfnamefont {P.}~\bibnamefont
		{Nozi\`{e}res}},\ }\href@noop {} {\bibfield  {journal} {\bibinfo  {journal}
		{J. Phys. Soc. Jpn.}\ }\textbf {\bibinfo {volume} {74}},\ \bibinfo {pages}
	{4} (\bibinfo {year} {2005})}\BibitemShut {NoStop}%
\bibitem [{\citenamefont {von Bergmann}\ \emph {et~al.}(2015)\citenamefont {von
		Bergmann}, \citenamefont {Ternes}, \citenamefont {Loth}, \citenamefont
	{Lutz},\ and\ \citenamefont {Heinrich}}]{Bergmann15}%
\BibitemOpen
\bibfield  {author} {\bibinfo {author} {\bibfnamefont {K.}~\bibnamefont {von
			Bergmann}}, \bibinfo {author} {\bibfnamefont {M.}~\bibnamefont {Ternes}},
	\bibinfo {author} {\bibfnamefont {S.}~\bibnamefont {Loth}}, \bibinfo {author}
	{\bibfnamefont {C.~P.}\ \bibnamefont {Lutz}}, \ and\ \bibinfo {author}
	{\bibfnamefont {A.~J.}\ \bibnamefont {Heinrich}},\ }\href@noop {} {\bibfield
	{journal} {\bibinfo  {journal} {Phys. Rev. Lett.}\ }\textbf {\bibinfo
		{volume} {114}},\ \bibinfo {pages} {076601} (\bibinfo {year}
	{2015})}\BibitemShut {NoStop}%
\bibitem [{\citenamefont {Bayat}\ \emph {et~al.}(2012)\citenamefont {Bayat},
	\citenamefont {Sodano},\ and\ \citenamefont {Bose}}]{Bayat12}%
\BibitemOpen
\bibfield  {author} {\bibinfo {author} {\bibfnamefont {A.}~\bibnamefont
		{Bayat}}, \bibinfo {author} {\bibfnamefont {P.}~\bibnamefont {Sodano}}, \
	and\ \bibinfo {author} {\bibfnamefont {S.}~\bibnamefont {Bose}},\ }\href@noop
{} {\bibfield  {journal} {\bibinfo  {journal} {Quant. Inform. Proc.}\
	}\textbf {\bibinfo {volume} {11}},\ \bibinfo {pages} {89} (\bibinfo {year}
	{2012})}\BibitemShut {NoStop}%
\bibitem [{\citenamefont {Anderson}(1961)}]{Anderson-1961}%
\BibitemOpen
\bibfield  {author} {\bibinfo {author} {\bibfnamefont {P.~W.}\ \bibnamefont
		{Anderson}},\ }\href {http://link.aps.org/doi/10.1103/PhysRev.124.41}
{\bibfield  {journal} {\bibinfo  {journal} {Phys. Rev.}\ }\textbf {\bibinfo
		{volume} {124}},\ \bibinfo {pages} {41} (\bibinfo {year} {1961})}\BibitemShut
{NoStop}%
\bibitem [{\citenamefont {Krishna-murthy}\ \emph {et~al.}(1980)\citenamefont
	{Krishna-murthy}, \citenamefont {Wilkins},\ and\ \citenamefont
	{Wilson}}]{Krishna-1980a}%
\BibitemOpen
\bibfield  {author} {\bibinfo {author} {\bibfnamefont {H.~R.}\ \bibnamefont
		{Krishna-murthy}}, \bibinfo {author} {\bibfnamefont {J.~W.}\ \bibnamefont
		{Wilkins}}, \ and\ \bibinfo {author} {\bibfnamefont {K.~G.}\ \bibnamefont
		{Wilson}},\ }\href {\doibase 10.1103/PhysRevB.21.1003} {\bibfield  {journal}
	{\bibinfo  {journal} {Phys. Rev. B}\ }\textbf {\bibinfo {volume} {21}},\
	\bibinfo {pages} {1003} (\bibinfo {year} {1980})}\BibitemShut {NoStop}%
\bibitem [{\citenamefont {Bulla}\ \emph {et~al.}(2008)\citenamefont {Bulla},
	\citenamefont {Costi},\ and\ \citenamefont {Pruschke}}]{Bulla-Rev-2008}%
\BibitemOpen
\bibfield  {author} {\bibinfo {author} {\bibfnamefont {R.}~\bibnamefont
		{Bulla}}, \bibinfo {author} {\bibfnamefont {T.~A.}\ \bibnamefont {Costi}}, \
	and\ \bibinfo {author} {\bibfnamefont {T.}~\bibnamefont {Pruschke}},\ }\href
{http://link.aps.org/doi/10.1103/RevModPhys.80.395} {\bibfield  {journal}
	{\bibinfo  {journal} {Rev. Mod. Phys.}\ }\textbf {\bibinfo {volume} {80}},\
	\bibinfo {pages} {395} (\bibinfo {year} {2008})}\BibitemShut {NoStop}%
\bibitem [{\citenamefont {Logan}\ \emph {et~al.}(2014)\citenamefont {Logan},
	\citenamefont {Tucker},\ and\ \citenamefont {Galpin}}]{Logan-2014}%
\BibitemOpen
\bibfield  {author} {\bibinfo {author} {\bibfnamefont {D.~E.}\ \bibnamefont
		{Logan}}, \bibinfo {author} {\bibfnamefont {A.~P.}\ \bibnamefont {Tucker}}, \
	and\ \bibinfo {author} {\bibfnamefont {M.~R.}\ \bibnamefont {Galpin}},\
}\href {http://link.aps.org/doi/10.1103/PhysRevB.90.075150} {\bibfield
{journal} {\bibinfo  {journal} {Phys. Rev. B}\ }\textbf {\bibinfo {volume}
	{90}},\ \bibinfo {pages} {075150} (\bibinfo {year} {2014})}\BibitemShut
{NoStop}%
\bibitem [{\citenamefont {{\v Z}itko}\ and\ \citenamefont
	{Horvat}(2016)}]{ZitkoAlen-2016}%
\BibitemOpen
\bibfield  {author} {\bibinfo {author} {\bibfnamefont {R.}~\bibnamefont {{\v
				Z}itko}}\ and\ \bibinfo {author} {\bibfnamefont {A.}~\bibnamefont {Horvat}},\
}\href {\doibase 10.1103/PhysRevB.94.125138} {\bibfield  {journal} {\bibinfo
	{journal} {Physical Review B}\ }\textbf {\bibinfo {volume} {94}},\ \bibinfo
{pages} {125138} (\bibinfo {year} {2016})}\BibitemShut {NoStop}%
\bibitem [{\citenamefont {Takegahara}\ \emph {et~al.}(1992)\citenamefont
	{Takegahara}, \citenamefont {Shimizu},\ and\ \citenamefont
	{Sakai}}]{Takegahara-1992}%
\BibitemOpen
\bibfield  {author} {\bibinfo {author} {\bibfnamefont {K.}~\bibnamefont
		{Takegahara}}, \bibinfo {author} {\bibfnamefont {Y.}~\bibnamefont {Shimizu}},
	\ and\ \bibinfo {author} {\bibfnamefont {O.}~\bibnamefont {Sakai}},\
}\bibfield  {booktitle} {\emph {\bibinfo {booktitle} {Journal of the Physical
		Society of Japan}},\ }\href {\doibase 10.1143/JPSJ.61.3443} {\bibfield
{journal} {\bibinfo  {journal} {Journal of the Physical Society of Japan}\
}\textbf {\bibinfo {volume} {61}},\ \bibinfo {pages} {3443} (\bibinfo {year}
{1992})}\BibitemShut {NoStop}%
\bibitem [{\citenamefont {Takegahara}\ \emph {et~al.}(1993)\citenamefont
	{Takegahara}, \citenamefont {Shimizu}, \citenamefont {Goto},\ and\
	\citenamefont {Sakai}}]{Takegahara-1993}%
\BibitemOpen
\bibfield  {author} {\bibinfo {author} {\bibfnamefont {K.}~\bibnamefont
		{Takegahara}}, \bibinfo {author} {\bibfnamefont {Y.}~\bibnamefont {Shimizu}},
	\bibinfo {author} {\bibfnamefont {N.}~\bibnamefont {Goto}}, \ and\ \bibinfo
	{author} {\bibfnamefont {O.}~\bibnamefont {Sakai}},\ }\href {\doibase
	https://doi.org/10.1016/0921-4526(93)90579-U} {\bibfield  {journal} {\bibinfo
		{journal} {Physica B: Condensed Matter}\ }\textbf {\bibinfo {volume}
		{186-188}},\ \bibinfo {pages} {381} (\bibinfo {year} {1993})}\BibitemShut
{NoStop}%
\bibitem [{\citenamefont {Saso}(1992)}]{Saso-1992}%
\BibitemOpen
\bibfield  {author} {\bibinfo {author} {\bibfnamefont {T.}~\bibnamefont
		{Saso}},\ }\bibfield  {booktitle} {\emph {\bibinfo {booktitle} {Journal of
			the Physical Society of Japan}},\ }\href {\doibase 10.1143/JPSJ.61.3439}
{\bibfield  {journal} {\bibinfo  {journal} {Journal of the Physical Society
			of Japan}\ }\textbf {\bibinfo {volume} {61}},\ \bibinfo {pages} {3439}
	(\bibinfo {year} {1992})}\BibitemShut {NoStop}%
\bibitem [{\citenamefont {Ogura}\ and\ \citenamefont
	{Saso}(1993)}]{Ogura-1993}%
\BibitemOpen
\bibfield  {author} {\bibinfo {author} {\bibfnamefont {J.}~\bibnamefont
		{Ogura}}\ and\ \bibinfo {author} {\bibfnamefont {T.}~\bibnamefont {Saso}},\
}\bibfield  {booktitle} {\emph {\bibinfo {booktitle} {Journal of the Physical
		Society of Japan}},\ }\href {\doibase 10.1143/JPSJ.62.4364} {\bibfield
{journal} {\bibinfo  {journal} {Journal of the Physical Society of Japan}\
}\textbf {\bibinfo {volume} {62}},\ \bibinfo {pages} {4364} (\bibinfo {year}
{1993})}\BibitemShut {NoStop}%
\bibitem [{\citenamefont {Cruz}\ \emph {et~al.}(1995)\citenamefont {Cruz},
	\citenamefont {Phillips},\ and\ \citenamefont {Neto}}]{Cruz-1995}%
\BibitemOpen
\bibfield  {author} {\bibinfo {author} {\bibfnamefont {L.}~\bibnamefont
		{Cruz}}, \bibinfo {author} {\bibfnamefont {P.}~\bibnamefont {Phillips}}, \
	and\ \bibinfo {author} {\bibfnamefont {A.~H.~C.}\ \bibnamefont {Neto}},\
}\bibfield  {booktitle} {\emph {\bibinfo {booktitle} {Europhysics Letters
		(EPL)}},\ }\href {\doibase 10.1209/0295-5075/29/5/007} {\ \textbf {\bibinfo
	{volume} {29}},\ \bibinfo {pages} {389} (\bibinfo {year} {1995})}\BibitemShut
{NoStop}%
\bibitem [{\citenamefont {Yu}\ and\ \citenamefont {Guerrero}(1996)}]{Yu-1996}%
\BibitemOpen
\bibfield  {author} {\bibinfo {author} {\bibfnamefont {C.~C.}\ \bibnamefont
		{Yu}}\ and\ \bibinfo {author} {\bibfnamefont {M.}~\bibnamefont {Guerrero}},\
}\href {\doibase 10.1103/PhysRevB.54.8556} {\bibfield  {journal} {\bibinfo
	{journal} {Physical Review B}\ }\textbf {\bibinfo {volume} {54}},\ \bibinfo
{pages} {8556} (\bibinfo {year} {1996})}\BibitemShut {NoStop}%
\bibitem [{\citenamefont {Chen}\ and\ \citenamefont
	{Jayaprakash}(1998)}]{Chen-1998}%
\BibitemOpen
\bibfield  {author} {\bibinfo {author} {\bibfnamefont {K.}~\bibnamefont
		{Chen}}\ and\ \bibinfo {author} {\bibfnamefont {C.}~\bibnamefont
		{Jayaprakash}},\ }\href {\doibase 10.1103/PhysRevB.57.5225} {\bibfield
	{journal} {\bibinfo  {journal} {Physical Review B}\ }\textbf {\bibinfo
		{volume} {57}},\ \bibinfo {pages} {5225} (\bibinfo {year}
	{1998})}\BibitemShut {NoStop}%
\bibitem [{\citenamefont {Galpin}\ and\ \citenamefont
	{Logan}(2008{\natexlab{a}})}]{Galpin-PRB2008}%
\BibitemOpen
\bibfield  {author} {\bibinfo {author} {\bibfnamefont {M.~R.}\ \bibnamefont
		{Galpin}}\ and\ \bibinfo {author} {\bibfnamefont {D.~E.}\ \bibnamefont
		{Logan}},\ }\href {\doibase 10.1103/PhysRevB.77.195108} {\bibfield  {journal}
	{\bibinfo  {journal} {Phys. Rev. B}\ }\textbf {\bibinfo {volume} {77}},\
	\bibinfo {pages} {195108} (\bibinfo {year} {2008}{\natexlab{a}})}\BibitemShut
{NoStop}%
\bibitem [{\citenamefont {Galpin}\ and\ \citenamefont
	{Logan}(2008{\natexlab{b}})}]{Galpin-EPJB2008}%
\BibitemOpen
\bibfield  {author} {\bibinfo {author} {\bibfnamefont {M.~R.}\ \bibnamefont
		{Galpin}}\ and\ \bibinfo {author} {\bibfnamefont {D.~E.}\ \bibnamefont
		{Logan}},\ }\href {\doibase 10.1140/epjb/e2008-00138-5} {\bibfield  {journal}
	{\bibinfo  {journal} {The European Physical Journal B}\ }\textbf {\bibinfo
		{volume} {62}},\ \bibinfo {pages} {129} (\bibinfo {year}
	{2008}{\natexlab{b}})}\BibitemShut {NoStop}%
\bibitem [{\citenamefont {Moca}\ and\ \citenamefont {Roman}(2010)}]{Moca-2010}%
\BibitemOpen
\bibfield  {author} {\bibinfo {author} {\bibfnamefont {C.~P.}\ \bibnamefont
		{Moca}}\ and\ \bibinfo {author} {\bibfnamefont {A.}~\bibnamefont {Roman}},\
}\href {\doibase 10.1103/PhysRevB.81.235106} {\bibfield  {journal} {\bibinfo
	{journal} {Phys. Rev. B}\ }\textbf {\bibinfo {volume} {81}},\ \bibinfo
{pages} {235106} (\bibinfo {year} {2010})}\BibitemShut {NoStop}%
\bibitem [{\citenamefont {Diniz}\ \emph {et~al.}(2020)\citenamefont {Diniz},
	\citenamefont {Diniz}, \citenamefont {Martins},\ and\ \citenamefont
	{Vernek}}]{Diniz-2020}%
\BibitemOpen
\bibfield  {author} {\bibinfo {author} {\bibfnamefont {G.}~\bibnamefont
		{Diniz}}, \bibinfo {author} {\bibfnamefont {G.~S.}\ \bibnamefont {Diniz}},
	\bibinfo {author} {\bibfnamefont {G.~B.}\ \bibnamefont {Martins}}, \ and\
	\bibinfo {author} {\bibfnamefont {E.}~\bibnamefont {Vernek}},\ }\href
{\doibase 10.1103/PhysRevB.101.125115} {\bibfield  {journal} {\bibinfo
		{journal} {Phys. Rev. B}\ }\textbf {\bibinfo {volume} {101}},\ \bibinfo
	{pages} {125115} (\bibinfo {year} {2020})}\BibitemShut {NoStop}%
\bibitem [{\citenamefont {Lenz}\ \emph {et~al.}(2013)\citenamefont {Lenz},
	\citenamefont {Urban},\ and\ \citenamefont {Bercioux}}]{Lenz-2013}%
\BibitemOpen
\bibfield  {author} {\bibinfo {author} {\bibfnamefont {L.}~\bibnamefont
		{Lenz}}, \bibinfo {author} {\bibfnamefont {D.~F.}\ \bibnamefont {Urban}}, \
	and\ \bibinfo {author} {\bibfnamefont {D.}~\bibnamefont {Bercioux}},\ }\href
{\doibase 10.1140/epjb/e2013-40760-4} {\bibfield  {journal} {\bibinfo
		{journal} {Eur. Phys. J. B}\ }\textbf {\bibinfo {volume} {86}} (\bibinfo
	{year} {2013}),\ 10.1140/epjb/e2013-40760-4}\BibitemShut {NoStop}%
\bibitem [{\citenamefont {Zalom}\ \emph {et~al.}(2021)\citenamefont {Zalom},
	\citenamefont {Pokorn{\'y}},\ and\ \citenamefont {Novotn{\'y}}}]{Zalom-2021}%
\BibitemOpen
\bibfield  {author} {\bibinfo {author} {\bibfnamefont {P.}~\bibnamefont
		{Zalom}}, \bibinfo {author} {\bibfnamefont {V.}~\bibnamefont {Pokorn{\'y}}},
	\ and\ \bibinfo {author} {\bibfnamefont {T.}~\bibnamefont {Novotn{\'y}}},\
}\href {\doibase 10.1103/PhysRevB.103.035419} {\bibfield  {journal} {\bibinfo
	{journal} {Physical Review B}\ }\textbf {\bibinfo {volume} {103}},\ \bibinfo
{pages} {035419} (\bibinfo {year} {2021})}\BibitemShut {NoStop}%
\bibitem [{\citenamefont {Doma{\'n}ski}\ \emph {et~al.}(2017)\citenamefont
	{Doma{\'n}ski}, \citenamefont {{\v Z}onda}, \citenamefont {Pokorn{\'y}},
	\citenamefont {G{\'o}rski}, \citenamefont {Jani{\v s}},\ and\ \citenamefont
	{Novotn{\'y}}}]{Domanski-2017}%
\BibitemOpen
\bibfield  {author} {\bibinfo {author} {\bibfnamefont {T.}~\bibnamefont
		{Doma{\'n}ski}}, \bibinfo {author} {\bibfnamefont {M.}~\bibnamefont {{\v
				Z}onda}}, \bibinfo {author} {\bibfnamefont {V.}~\bibnamefont {Pokorn{\'y}}},
	\bibinfo {author} {\bibfnamefont {G.}~\bibnamefont {G{\'o}rski}}, \bibinfo
	{author} {\bibfnamefont {V.}~\bibnamefont {Jani{\v s}}}, \ and\ \bibinfo
	{author} {\bibfnamefont {T.}~\bibnamefont {Novotn{\'y}}},\ }\href
{http://link.aps.org/doi/10.1103/PhysRevB.95.045104} {\bibfield  {journal}
	{\bibinfo  {journal} {Physical Review B}\ }\textbf {\bibinfo {volume} {95}},\
	\bibinfo {pages} {045104} (\bibinfo {year} {2017})}\BibitemShut {NoStop}%
\bibitem [{Note1()}]{Note1}%
\BibitemOpen
\bibinfo {note} {$\omega ^+$ emphasizes to which part of the complex
	$z$-plain the function belongs, i.e.~above the real axis of $z$.}\BibitemShut
{Stop}%
\bibitem [{\citenamefont {Kadlecov{\'a}}\ \emph {et~al.}(2017)\citenamefont
	{Kadlecov{\'a}}, \citenamefont {{\v Z}onda},\ and\ \citenamefont
	{Novotn{\'y}}}]{Kadlecova-2017}%
\BibitemOpen
\bibfield  {author} {\bibinfo {author} {\bibfnamefont {A.}~\bibnamefont
		{Kadlecov{\'a}}}, \bibinfo {author} {\bibfnamefont {M.}~\bibnamefont {{\v
				Z}onda}}, \ and\ \bibinfo {author} {\bibfnamefont {T.}~\bibnamefont
		{Novotn{\'y}}},\ }\href {\doibase 10.1103/PhysRevB.95.195114} {\bibfield
	{journal} {\bibinfo  {journal} {Phys. Rev. B}\ }\textbf {\bibinfo {volume}
		{95}},\ \bibinfo {pages} {195114} (\bibinfo {year} {2017})}\BibitemShut
{NoStop}%
\bibitem [{\citenamefont {Bulla}\ \emph {et~al.}(1994)\citenamefont {Bulla},
	\citenamefont {Keller},\ and\ \citenamefont {Pruschke}}]{Bulla-1994}%
\BibitemOpen
\bibfield  {author} {\bibinfo {author} {\bibfnamefont {R.}~\bibnamefont
		{Bulla}}, \bibinfo {author} {\bibfnamefont {J.}~\bibnamefont {Keller}}, \
	and\ \bibinfo {author} {\bibfnamefont {T.}~\bibnamefont {Pruschke}},\ }\href
{\doibase 10.1007/BF01307671} {\bibfield  {journal} {\bibinfo  {journal} {Z.
			Phys. B Condens. Matter}\ }\textbf {\bibinfo {volume} {94}},\ \bibinfo
	{pages} {195} (\bibinfo {year} {1994})}\BibitemShut {NoStop}%
\bibitem [{\citenamefont {{\v Z}itko}(2014)}]{Ljubljana-code}%
\BibitemOpen
\bibfield  {author} {\bibinfo {author} {\bibfnamefont {R.}~\bibnamefont {{\v
				Z}itko}},\ }\href@noop {} {\enquote {\bibinfo {title} {{NRG L}jubljana - open
			source numerical renormalization group code},}\ } (\bibinfo {year} {2014}),\
\bibinfo {note} {nrgljubljana.ijs.si}\BibitemShut {NoStop}%
\bibitem [{\citenamefont {{\v Z}itko}\ and\ \citenamefont
	{Pruschke}(2009)}]{ZitkoPruschke-2009}%
\BibitemOpen
\bibfield  {author} {\bibinfo {author} {\bibfnamefont {R.}~\bibnamefont {{\v
				Z}itko}}\ and\ \bibinfo {author} {\bibfnamefont {T.}~\bibnamefont
		{Pruschke}},\ }\href {\doibase 10.1103/PhysRevB.79.085106} {\bibfield
	{journal} {\bibinfo  {journal} {Phys. Rev. B}\ }\textbf {\bibinfo {volume}
		{79}},\ \bibinfo {pages} {085106} (\bibinfo {year} {2009})}\BibitemShut
{NoStop}%
\bibitem [{\citenamefont {Bulla}\ \emph {et~al.}(1998)\citenamefont {Bulla},
	\citenamefont {Hewson},\ and\ \citenamefont {Pruschke}}]{Bulla-1998}%
\BibitemOpen
\bibfield  {author} {\bibinfo {author} {\bibfnamefont {R.}~\bibnamefont
		{Bulla}}, \bibinfo {author} {\bibfnamefont {A.~C.}\ \bibnamefont {Hewson}}, \
	and\ \bibinfo {author} {\bibfnamefont {T.}~\bibnamefont {Pruschke}},\ }\href
{\doibase 10.1088/0953-8984/10/37/021} {\bibfield  {journal} {\bibinfo
		{journal} {Journal of Physics: Condensed Matter}\ }\textbf {\bibinfo {volume}
		{10}},\ \bibinfo {pages} {8365} (\bibinfo {year} {1998})}\BibitemShut
{NoStop}%
\bibitem [{\citenamefont {Hewson}(1993)}]{Hewson-1993}%
\BibitemOpen
\bibfield  {author} {\bibinfo {author} {\bibfnamefont {A.~C.}\ \bibnamefont
		{Hewson}},\ }\href {\doibase 10.1017/CBO9780511470752} {\emph {\bibinfo
		{title} {The Kondo Problem to Heavy Fermions}}},\ Cambridge Studies in
Magnetism\ (\bibinfo  {publisher} {Cambridge University Press},\ \bibinfo
{year} {1993})\BibitemShut {NoStop}%
\bibitem [{\citenamefont {{\v Z}itko}(2007)}]{Zitko-PhDthesis}%
\BibitemOpen
\bibfield  {author} {\bibinfo {author} {\bibfnamefont {R.}~\bibnamefont {{\v
				Z}itko}},\ }\emph {\bibinfo {title} {Many-particle effects in resonant
		tunneling of electrons through nanostructures}},\ \href@noop {} {Ph.D.
	thesis},\ \bibinfo  {school} {University of Ljubljana} (\bibinfo {year}
{2007})\BibitemShut {NoStop}%
\bibitem [{\citenamefont {Yang}\ and\ \citenamefont {Tong}(2020)}]{Yang-2020}%
\BibitemOpen
\bibfield  {author} {\bibinfo {author} {\bibfnamefont {K.}~\bibnamefont
		{Yang}}\ and\ \bibinfo {author} {\bibfnamefont {N.-H.}\ \bibnamefont
		{Tong}},\ }\href {\doibase 10.1103/PhysRevB.102.085125} {\bibfield  {journal}
	{\bibinfo  {journal} {Physical Review B}\ }\textbf {\bibinfo {volume}
		{102}},\ \bibinfo {pages} {085125} (\bibinfo {year} {2020})}\BibitemShut
{NoStop}%
\bibitem [{\citenamefont {Satori}\ \emph {et~al.}(1992)\citenamefont {Satori},
	\citenamefont {Shiba}, \citenamefont {Sakai},\ and\ \citenamefont
	{Shimizu}}]{Satori-1992}%
\BibitemOpen
\bibfield  {author} {\bibinfo {author} {\bibfnamefont {K.}~\bibnamefont
		{Satori}}, \bibinfo {author} {\bibfnamefont {H.}~\bibnamefont {Shiba}},
	\bibinfo {author} {\bibfnamefont {O.}~\bibnamefont {Sakai}}, \ and\ \bibinfo
	{author} {\bibfnamefont {Y.}~\bibnamefont {Shimizu}},\ }\href {\doibase
	10.1143/JPSJ.61.3239} {\bibfield  {journal} {\bibinfo  {journal} {J. Phys.
			Soc. Japan.}\ }\textbf {\bibinfo {volume} {61}},\ \bibinfo {pages} {3239}
	(\bibinfo {year} {1992})}\BibitemShut {NoStop}%
\bibitem [{\citenamefont {Bruus}\ and\ \citenamefont
	{Flensberg}(2004)}]{Flensberg-book}%
\BibitemOpen
\bibfield  {author} {\bibinfo {author} {\bibfnamefont {H.}~\bibnamefont
		{Bruus}}\ and\ \bibinfo {author} {\bibfnamefont {K.}~\bibnamefont
		{Flensberg}},\ }\href@noop {} {\emph {\bibinfo {title} {Many-body Quantum
			Theory in Condensed Matter Physics: an introduction}}}\ (\bibinfo
{publisher} {Oxford University Press},\ \bibinfo {year} {2004})\BibitemShut
{NoStop}%
\bibitem [{\citenamefont {Jellinggaard}\ \emph {et~al.}(2016)\citenamefont
	{Jellinggaard}, \citenamefont {Grove-Rasmussen}, \citenamefont {Madsen},\
	and\ \citenamefont {Nyg{\aa}rd}}]{Jellinggaard-2016}%
\BibitemOpen
\bibfield  {author} {\bibinfo {author} {\bibfnamefont {A.}~\bibnamefont
		{Jellinggaard}}, \bibinfo {author} {\bibfnamefont {K.}~\bibnamefont
		{Grove-Rasmussen}}, \bibinfo {author} {\bibfnamefont {M.~H.}\ \bibnamefont
		{Madsen}}, \ and\ \bibinfo {author} {\bibfnamefont {J.}~\bibnamefont
		{Nyg{\aa}rd}},\ }\href {http://link.aps.org/doi/10.1103/PhysRevB.94.064520}
{\bibfield  {journal} {\bibinfo  {journal} {Phys. Rev. B}\ }\textbf {\bibinfo
		{volume} {94}},\ \bibinfo {pages} {064520} (\bibinfo {year}
	{2016})}\BibitemShut {NoStop}%
\bibitem [{\citenamefont {Whiticar}\ \emph {et~al.}(2021)\citenamefont
	{Whiticar}, \citenamefont {Fornieri}, \citenamefont {Banerjee}, \citenamefont
	{Drachmann}, \citenamefont {Gronin}, \citenamefont {Gardner}, \citenamefont
	{Lindemann}, \citenamefont {Manfra},\ and\ \citenamefont
	{Marcus}}]{Whiticar-2021}%
\BibitemOpen
\bibfield  {author} {\bibinfo {author} {\bibfnamefont {A.~M.}\ \bibnamefont
		{Whiticar}}, \bibinfo {author} {\bibfnamefont {A.}~\bibnamefont {Fornieri}},
	\bibinfo {author} {\bibfnamefont {A.}~\bibnamefont {Banerjee}}, \bibinfo
	{author} {\bibfnamefont {A.~C.~C.}\ \bibnamefont {Drachmann}}, \bibinfo
	{author} {\bibfnamefont {S.}~\bibnamefont {Gronin}}, \bibinfo {author}
	{\bibfnamefont {G.~C.}\ \bibnamefont {Gardner}}, \bibinfo {author}
	{\bibfnamefont {T.}~\bibnamefont {Lindemann}}, \bibinfo {author}
	{\bibfnamefont {M.~J.}\ \bibnamefont {Manfra}}, \ and\ \bibinfo {author}
	{\bibfnamefont {C.~M.}\ \bibnamefont {Marcus}},\ }\href@noop {} {\enquote
	{\bibinfo {title} {Zeeman-driven parity transitions in an andreev quantum
			dot},}\ } (\bibinfo {year} {2021}),\ \Eprint
{http://arxiv.org/abs/2101.09706} {arXiv:2101.09706 [cond-mat.mes-hall]}
\BibitemShut {NoStop}%
\end{thebibliography}

%

\end{document}